\documentclass[3p]{elsarticle}

\usepackage{lineno,hyperref}

\usepackage[labelfont=bf]{caption}
\usepackage{caption}
\DeclareCaptionLabelSeparator{point}{.~}
\usepackage{amsmath}
\usepackage{algorithm}
\usepackage[noend]{algpseudocode}
\usepackage{booktabs}
\usepackage[skip=0.1\baselineskip]{caption}
\usepackage{subcaption}
\usepackage{multirow}
\usepackage[table,xcdraw]{xcolor}
\usepackage{tabu}
\usepackage{xcolor}
\usepackage{graphicx}
\usepackage{nomencl}
\usepackage{multicol}
\usepackage{etoolbox}

\journal{arXiv}

\begin{document}
\begin{frontmatter}

\title{Band Relevance Factor (BRF): a novel automatic frequency band selection method based on vibration analysis for rotating machinery}

\author[mymainaddress]{Lucas Costa Brito\corref{mycorrespondingauthor}}
\cortext[mycorrespondingauthor]{Corresponding author}
\ead{lucas.brito@ufu.br | brito.lcb@gmail.com}
\author[mysecondaryaddress]{Gian Antonio Susto}
\author[mythirdaryaddress]{Jorge Nei Brito}
\author[mymainaddress]{Marcus Antonio Viana Duarte}

\address[mymainaddress]{School of Mechanical Engineering, Federal University of Uberlândia, Av. João N. Ávila, 2121, Uberlândia, Brazil}
\address[mysecondaryaddress]{Department of Information Engineering, University of Padova, Via Gradenigo 6/B, 35131, Padova, Italy}
\address[mythirdaryaddress]{Department of Mechanical Engineering, Federal University of São João del Rei, P.Orlando, 170, São João del Rei, Brazil}

\begin{abstract}
The monitoring of rotating machinery has now become a fundamental activity in the industry, given the high criticality in production processes. Extracting useful information from relevant signals is a key factor for effective monitoring: studies in the areas of Informative Frequency Band selection (IFB) and Feature Extraction/Selection have demonstrated to be effective approaches. However, in general, 
Typical methods in such areas focuses on identifying bands where impulsive excitations are present or on analyzing the relevance of the features after its signal extraction: both approaches lack in terms of procedure automation and efficiency. Typically, the approaches presented in the literature fail to identify frequencies relevant for the vibration analysis of a rotating machinery; moreover, with such approaches features can be extracted from irrelevant bands, leading to additional complexity in the analysis. To overcome such problems, the present study proposes a new approach called Band Relevance Factor (BRF). BRF aims to perform an automatic selection of all relevant frequency bands for a vibration analysis of a rotating machine based on spectral entropy. The results are presented through a relevance ranking and can be visually analyzed through a heatmap. The effectiveness of the approach is validated in a synthetically created dataset and two real dataset, showing that the BRF is able to identify the bands that present relevant information for the analysis of rotating machinery.
\newline
\end{abstract}

\begin{keyword}
\texttt Band Relevance Factor \sep Informative Frequency Band Selection \sep Feature Extraction \sep Feature Selection \sep Rotating Machinery \sep Spectral Entropy 
\end{keyword}
\end{frontmatter}


\section{Introduction}

Rotating machinery is one of the most widely used mechanical equipment of modern industry \cite{E*}. Due to its criticality in the processes, several techniques are used to monitor the integrity of each asset, among which the vibration analysis stands out \cite{xai_lucas}.

The vibration signal is composed of characteristics that refer to the behavior of the system under analysis \cite{I1}. Vibration signals from rotating machinery usually present frequency bands where the dynamic behavior of the equipment and certain faults can be analyzed. Determining the regions that present relevant information is essential to assist the specialist in identifying defects, evaluating the quality of the collected signal, allowing the extraction of significant features for use in statistical and machine learning (ML) models, or even helping to explain the artificial intelligence models, an area of recent interest called: Explainable Artificial Intelligence (XAI).

Despite the amount of information that can be obtained by analyzing the signals, the collected vibration signals of rotating machinery are usually weakened and disturbed by the strong environment noises and other neighboring components \cite{E*}. This makes it necessary to develop methods to assist in extracting relevant information.

Two major areas stand out in the study of identifying relevant information in signals: i) Feature Extraction/Selection ii) Informative Frequency Band selection (IFB).
Feature extraction/selection is one of the fundamental areas of study for signal analysis and applications of statistical and machine learning models. Performing the extraction and selection of relevant features allows knowing the signal under study and avoiding the introduction of non-significant or redundant features that tend to reduce the assertiveness of the model \cite{infomartics_brito}. For the study of rotating machinery, as shown in \cite{xai_lucas}, the features to detect faults and to analyze the dynamic behavior of rotating machinery using vibration signals are commonly extracted from time, frequency and time-frequency domain.

Correctly identifying the relevant frequency bands for feature extraction/selection is critical to successful application. When features are extracted directly from the raw signal without a previous relevance analysis, there is a high chance of introducing a lot of irrelevant information into the analysis. As presented by \cite{Feat.XX} if the extracted features are informative enough, any classification algorithm may perform very well. Unfortunately, it is a really challenging task that is almost impossible to fulfill in real world applications.

Studies have been developed, and can be divided into three categories, i.e., filters, wrappers, and embedded methods. \cite{Feat.1} propose a fault diagnosis strategy based on improved multiscale dispersion entropy (IMDE) and max-relevance min-redundancy (mRMR). The analysis show that the proposed method can extract effectively fault information. \cite{Feat.2} use the two-stage feature selection, where Relief is applied for preliminarily selection and in the reselection process, Binary Particle Swarm Optimization (BPSO). \cite{Feat.3} proposes an end-to-end feature selection and diagnosis method for rotating machinery combining with dimensionality reduction. \cite{LI2017295} analyzed the multi-scale symbolic dynamic entropy and mRMR feature selection for fault detection in planetary gearboxes. Other works can be found at \cite{Canedo}, where a general review on the topic is presented. Also in \cite{Feat.XX} where an analysis of the importance of extracting features is performed.

In most of the works presented in the literature, the main focus is to obtain relevant information from the extracted features, which, combined with artificial intelligence methods, may be capable of diagnosing the operating condition of the equipment. Despite the good results presented, it is noted that the choice of frequency bands still needs to be automated, since, in general, the methods proposed in the literate use the raw signals or equally divided bands, without a prior analysis of relevance. An additional scenario for the methods proposed in the literature is that, when the relevance is analyzed, it is related to the feature already extracted from the signals. In engineering applications, however, the machine users typically are looking for automatic methods to shorten the maintenance cycle and improve the diagnosis accuracy \cite{LEI2020106587}.

Another related area of study is called: Informative Frequency Band \cite{Accur.1} (some other terminologies are: Frequency Band Selection (FBS) \cite{accugram}, Optimal Band Selection \cite{Int.26,Accur.3}). Unlike feature extraction, the area mainly aims to study methods to distinguish regions where impulsive excitations occur, with such excitations being caused by faults like bearings and gear-related ones; given the difficulty of identifying the bands excited by impulses, in noisy signals, or in incipient defects, Informative Frequency Band is particularly appealing for researchers and practitioners.

\cite{Int.1} present a practical framework of Gini Index (GI) in the application of rotating machinery, comparing state-of-the-art and new methods, such as spectral kurtosis-based methods, decomposition methods, deconvolution methods: Kurtogram \cite{Int.24, Int.25}, Protrugram \cite{Int.26}, Autogram \cite{Int.30}, GI derivations. Such methods have however several limitations: for example, using the Envelope spectrum kurtosis-based method (Protrugram), it is not possible to analyze the impulsive excitations in the presence of harmonic interference. As for the Kurtosis-based method (minimum entropy deconvolution (MED) \cite{Int.40} and Kurtogram) it is also not possible to perform analysis in the presence of random impulse. \cite{Int.2} present a novel approach to detecting cyclic impulses in the presence of non-cyclic impulses, using conditional variance based (CVB) statistic/selector and compare the result to the state-of-arts methods: spectral kurtosis l2/l1 norm \cite{Int.2.25}, Alpha selector \cite{Int.2.17}, kurtogram, spectral Gini index \cite{Int.2.26}, spectral smoothness index \cite{Int.2.27,Int.2.28}, and the infogram \cite{Int.2.30}. In summary, for a signal with only Gaussian noise, no technique of the aforementioned indicated any band, as expected. For the cyclic and non-cyclic impulsive signal with Gaussian noise background, all methods showed good results. For the case in which the presence of the non-cyclic impulsive signal was much greater in relation to the cyclic impulse, the CVB selector was able to identify the frequency band corresponding to the cyclic impulses.

As can be seen, IFB works are focused on determining frequency regions where impulsive excitations are present, so that the signal can be filtered and a signal analysis technique such as envelope applied. The studies are extremely relevant, in view of the great difficulty in identifying such frequency bands, due to the low amplitudes presented when the defect is in the incipient stage. On the other hand, other frequencies present in the signal are also related to faults and/or dynamic behavior of the machine, which, because they present behavior different from the impulsive (eg, cyclic/harmonic/random) are not addressed by the aforementioned methods. Such frequency bands are fundamental for a vibration analysis performed by a human expert, and also for the automatic extraction of features for AI models. This is because it contains information regarding the dynamic behavior of the machine (e.g., rotation frequency), and various faults such as: unbalance, mechanical backlash, misalignment, etc.

Because most of the time, they are evident in the signal, due to their greater amplitude, such frequencies are not the focus of the works on IFB. However, aiming to support the human expert in decision making and Artificial Intelligence (AI) frameworks in the automatic extraction of relevant features and XAI, developing a method that can automatically identify such relevant frequency bands in the signal, contributes to the monitoring studies of rotating machinery and possible industrial applications.

Thus, the study proposes a new approach called BRF (Band Relevance Factor). The approach aims to perform an automatic selection of all relevant frequency bands for a vibration analysis of a rotating machinery based on spectral entropy.

Initially, the method allows the automatic evaluation of the quality of the collected signal, assessing whether it only presents noise, or if there is the presence of characteristic frequencies of the equipment. Furthermore, the method automatically points the vibration analyst to all relevant frequency bands in the signal. Finally, the use of the method as a pre-feature extraction stage allows the extraction of features in automatically selected relevant bands, for use in statistical and machine learning models.

Because it has a strong relationship with the behavior of rotating machinery, since the vast majority of faults are related to harmonic components and their multiples, spectral entropy was used. Due to its behavior, a signal from a healthy machine will show larger entropy value due to its high irregularity, while a faulted machine will have low entropy due to its low irregularity caused by the localized damage \cite{22*, LI2017295}, which allows mapping the most relevant bands. In addition to entropy, the approach combines the use of the root-mean-square value (rms), a parameter widely used in vibration analysis of rotating machinery, as it provides an overview of the energy present in the signal.

In summary, the main contributions of this paper are: i) A novel framework is proposed for automatic identification of relevant frequency bands in vibration signals; ii) A relevance ranking is proposed, to define among the selected bands the most important ones; iii) The heatmap is proposed to facilitate the visual analysis of the relevance ranking; iv) Possibility of applying the method in different rotating machinery and faults; v) Industrial application.

Due to the main characteristics, the work can be considered a contribution to two major areas: feature extraction/selection and IFB. Initially it shares the main objective of feature extraction, which is to extract relevant information from the signal, adding an automatic framework. On the other hand, it uses informative frequency band analysis more broadly (not focused only on impulsive defects), to automatically determine the relevant frequency bands in the signals. Such bands can be evaluated by the specialist during the analysis, or selected by an ML framework to evaluate the dynamic behavior of rotating machinery.

The remainder of this paper starts with a brief explanation about Entropy. The proposed method is presented in Section 3. Experimental procedure is shown in Section 4. Results and discussion are given in Section 5. Finally, Section 6 concludes this paper.

\section{Background}

\subsection{Entropy} \label{sec:ad}

Entropy, as a statistical measure, is able to quantify and detect changes in time series taking into account their nonlinear behavior \cite{E*}. Due to its characteristics, recently several entropy-based methods have been studied for rotating machinery. \cite{15*E,17*E} proposed the multi-scale permutation entropy for fault diagnosis in planetary gearboxes and rolling bearing. \cite{16*E} studied a fuzzy entropy approach for fault detection in bearing. In general, studies show promising results for the use of techniques in monitoring rotating machinery. \cite{53*E,54*E,56*E} presented studies involving Spectral Entropy in the detection of faults in rotating machines. Other studies can be found in \cite{E*,Entropy2R}.

The definition of entropy is proposed by Shannon to evaluate the irregularity and self-similarity of time series in information theory \cite{37.S}. For the discrete data series \emph{\{x1, x2,..., xn\}}, the  Shannon entropy H is defined as follows.

\begin{flalign}
    & H(x) = -\sum_{i=1}^{n} p(x_i)log(p(x_i)) & \label{eq:eq1}
\end{flalign}

where \emph{p} represents the probability of the time series {\{$x\textsubscript{i}$\}}

A bigger entropy indicates a more uncertainty or irregularity of time series and if a probability can be divided into the sum of several individual values, so does the Shannon entropy \cite{E*}. For a given time series, if the probability values of different states are similar, it is difficult to determine the future status, thus the time series has its maximum entropy value. In contrast, if there is only one state, the time series has its minimum entropy \cite{26.3}.

Among the different methods of using Shannon's Entropy proposed in the fault diagnosis of rotating machinery \cite{E*}, Spectral Entropy stands out.
The spectral entropy of a signal is a measure of its spectral distribution, and it is a normalized form of Shannon entropy, calculated as follow:

\begin{flalign}
    & S(x) = -\frac{\sum_{i=1}^{n} P_x log(P_x)}{log(n)} & \label{eq:eq2}
\end{flalign}

where, {$P\textsubscript{x}$} is the probability distribution and \emph{n} is the total frequency points used to normalized between 0 and 1. {$P\textsubscript{x}$} is described as follow:

\begin{flalign}
    & P_x = \frac{E_i}{\sum_{i}^{n} E_i} & \label{eq:eq3}
\end{flalign}

{$E\textsubscript{i}$} is the energy in each frequency \emph{i}. 

When the distribution of values is flat with equal or close amplitudes for each frequency component, Spectral entropy will result in high values, close to 1. Distribution similar to a signal with only noise. On the other hand, if the amplitudes are concentrated in a few frequency components, especially if only a few frequencies have non-zero amplitudes, the spectral entropy value will be close to 0 \cite{E1,E2}.

It is known that the vast majority of faults in rotating machines are related to harmonic components and their multiples. For this reason, it is possible to use entropy to analyze the signals, knowing that the vibration signal collected from a healthy machine has a larger entropy value due to its high irregularity, while that collected from a faulty rotating machinery has a smaller entropy value due to its low irregularity caused by the localized damage \cite{22*, LI2017295}.

\section{Proposed approach: Band Relevance Factor}

\begin{figure*}[ht]
  \centering
  \includegraphics[scale=0.75]{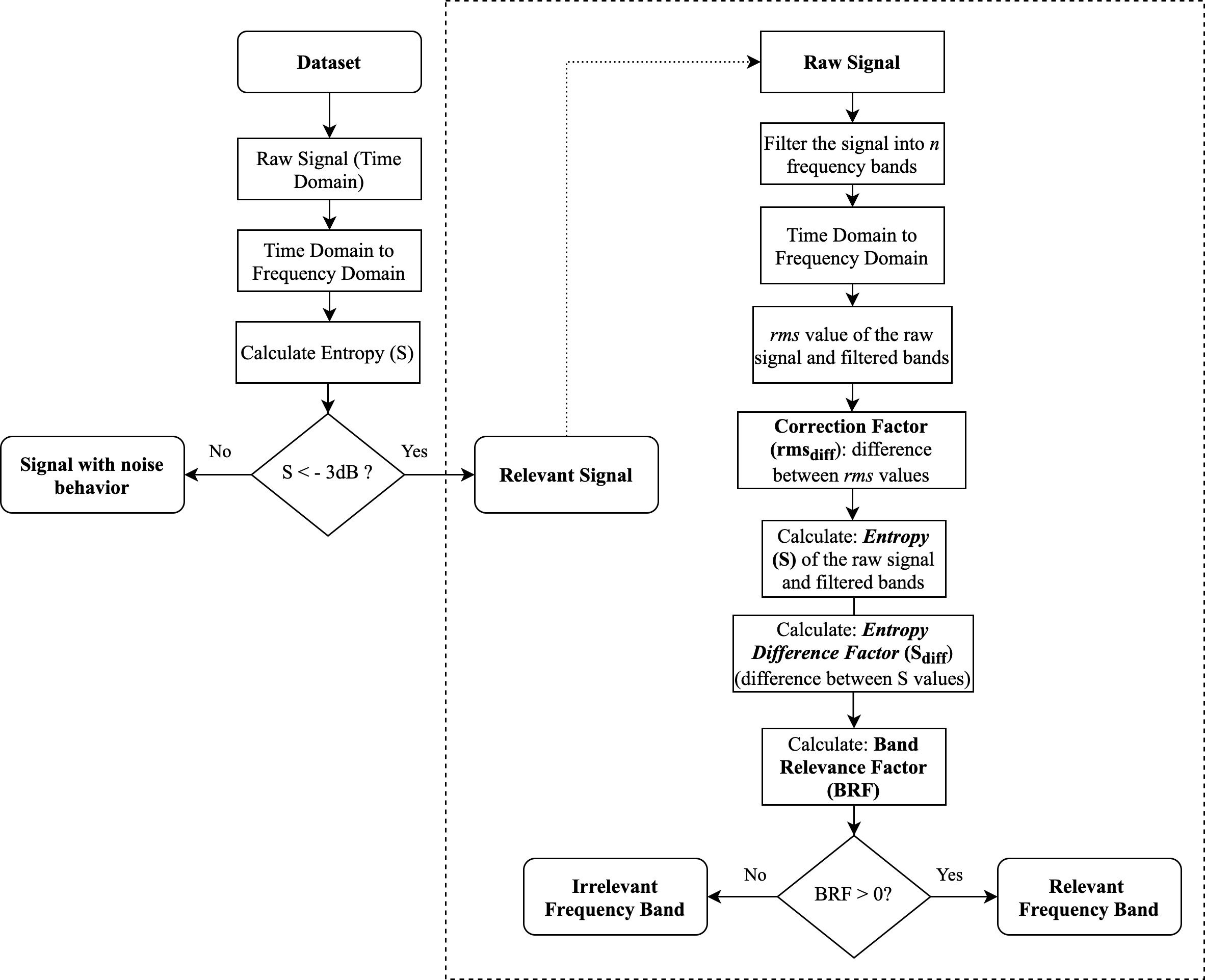}
  \caption{General framework of the proposed methodology.}
  \label{fig:Cap_BRF}
\end{figure*}

The proposed approach is presented in Fig. \ref{fig:Cap_BRF}.

Initially the original signal is transformed from the time domain to frequency domain and its entropy calculated. Signals with entropy greater than or equal to -3 decibels (dB) are considered to be composed only of noise and, therefore, irrelevant for the analysis. On the other hand, signals with entropy less than -3 dB are considered relevant, and likely to identify important frequency bands.

After being identified as relevant, the original signal is filtered in \textit{k}-level (bands of interest), in order to verify which bands are relevant in relation to the complete signal (being the amount of level defined by the user). Subsequently, the signal is transformed from time to frequency domain. The rms value of the filtered and complete signal is calculated, and the difference between them is called the Correction Factor ({$rms\textsubscript{diff}$}). The entropy of the complete signal and each band is calculated (\textit{S}). The difference between the entropy of the complete signal and each band is calculated, called the Entropy Difference Factor ({$S\textsubscript{diff}$}). Finally, the division between the Entropy Difference Factor and the Correction Factor is calculated to obtain the new proposed feature, called Band Relevance Factor (BRF). Positive values indicate that the band under analysis is relevant to the original signal, and negative values that the band is irrelevant to the original signal.

For the methodology, a so-called relevant signal is a signal that presents deterministic frequencies distinguishable from the background noise, without the need to perform advanced signal processing to remove it. Likewise, relevant frequency bands are regions that present relevant information for analysis and that can be identified through a spectral analysis, not involving advanced processing techniques. In addition, they allow the specialist to carry out an analysis regarding its characteristic, in order to understand the dynamic behavior of a system and/or identify related faults.

\subsection{Raw Signal Entropy}

The first step of the methodology consists of calculating the entropy of the raw (original) signal in order to measure the irregularity of the system. If the entropy value in dB is greater than or equal to -3 dB, the signal is considered noise. In this case no band is relevant in relation to the others, and therefore the following procedure does not need to be applied and the signal is not analyzed.

The -3 dB point is commonly used with filters. It indicates the frequency at which the associated power drops to half of its 50\% maximum value. A regular/noise (equiprobable) signal in the frequency domain tends to have roughly the same amplitude value at all frequencies. Therefore, when calculating the value of Spectral Entropy (\textit{S}), regular signals (equiprobable) will have \textit{S} equal to or close to 1, and ordered signals (harmonics) will have \textit{S} less than 1 and closer to 0. Analyzing the entropy value (assuming S = 1) for regular dB signals, it is known that S = 10{$log\textsubscript{10}$}(1) is equal to 0.

Thus, the value for regular/noise signals (equiprobable) analyzed on the dB scale will present values of entropy zero or close to zero. Assuming that some signals may be combinations of harmonics and noise, it is necessary to define a maximum entropy value from which it is possible to analyze the signals for fault detection (i.e., the signals are not purely noise). Assuming that 3 dB means a gain or reduction of 50\% of power in the associated frequency, and approximately 0 dB being the entropy value for a regular signal/noise, it is defined that for the entropy value of - 3dB, the system is proportionately more ordered, with less uncertainty and less variability. Being possible to separate the deterministic frequencies from the noise, and consequently perform the analysis of the behavior of the machine, and identify faults.

Therefore, if the \textit{S} of the original signal is less than -3 dB, there is the presence of relevant frequencies in the signal, which may be related to some fault characteristic or the dynamic behavior of the equipment, and therefore it is possible to continue the analysis. For higher \textit{S} values the signal is considered irrelevant.

\subsection{Correction Factor}

After analyzing the relevance of the complete signal, filtering the signal in \textit{k}-level (bands of interest) and transforming it to the frequency domain, the Correction Factor, proposed as a weighting in the BRF, is calculated as follow:

\begin{flalign}
    & rms_{diff} = rms_{base} - rms_{filtered} & \label{eq:eq4}
\end{flalign}

Where {$rms\textsubscript{base}$} is a statistical measure defined as the root mean square value of the original signal and {$rms\textsubscript{filtered}$} the root mean square value of the filtered band, both in dB. \textit{rms} measures the overall value of energy present in the signal, the greater the magnitude, the greater the \textit{rms} value from the global point of view.

A signal with many spectral components (e.g., excitation of resonances in bearings, mechanical looseness, cavitation etc.), will result in a larger entropy value due to its high irregularity, which may have amplitudes low or high. Consequently, if the BRF used only entropy as an analysis parameter, the ranking of the most relevant bands in the signal could present an error, since the amplitude would not be directly taken into account. Therefore, the Correction Factor is proposed to quantify the importance of the amplitude present in the frequency band through the difference between the \textit{rms} value of the original and filtered signal.

\subsection{Entropy Difference Factor}

The Entropy Difference Factor is proposed to evaluate the order, uncertainty and variability of the signal, calculated as:

\begin{flalign}
    & S_{diff} = 3 + S_{base} - S_{filtered} & \label{eq:eq5}
\end{flalign}

To provide decision making based on the {$S\textsubscript{diff}$} (+ or -), 3 dB was added to the {$S\textsubscript{diff}$} value. This allows to replace the decision threshold which was previously -3 dB to 0 dB. Thus, if the {$S\textsubscript{diff}$} difference is positive or equal to zero, the band under analysis is relevant, otherwise the band is not relevant to the analysis.

In the new scale, the value for regular/noise bands (equiprobable) analyzed in dB will present positive entropy values. Considering that the signal was approved in the first test, the value of {$S\textsubscript{base}$} has a negative value, so when applied to the equation we will have: (-) - (+) = -, therefore irrelevant for the signal.

In a signal with only one harmonic, the {$S\textsubscript{filtered}$} value of the band in which the harmonic is present will be less than or approximately the same as the full signal, resulting in a {$S\textsubscript{diff}$} equal to or approximately zero. In a signal with harmonic and noise, when filtering in the harmonic region, entropy tends to reduce, as it is a more ordered region (since the harmonic is present), and less uncertainty, due to the noise limitation, in this way, the {$S\textsubscript{filtered}$} entropy will be lower, ensuring a positive value.

\subsection{Band Relevance Factor (BRF)}

Although entropy quantifies the regularity of the signal, that is, the presence of harmonics or random frequencies, not all component failures present harmonic behavior, which can lead to an error in the identification ranking of the relevance of the band if only the Entropy Difference Factor is used. To overcome this problem, the Entropy Difference Factor is divided by the Correction Factor, resulting in the Band Relevance Factor (BRF).

The more energy in the band, the lower the value of the Correction Factor, and consequently the higher the value of the BRF, since it is in the denominator. In this way, it is possible to determine, among the selected bands, which is the one with the greatest relevance.

The Band Relevance Factor is obtained through the equation:

\begin{flalign}
    & BRF = \frac{S_{diff}}{rms_{diff}} & \label{eq:eq6}
\end{flalign}

Being, {$S\textsubscript{diff}$} the Entropy Difference Factor and {$rms\textsubscript{diff}$} the Correction Factor.

Positive values indicate that the band under analysis presents relevant information for analysis. In addition, the higher the BRF value, the more information that band presents for analysis, thus making it possible not only to define whether the band is relevant or not, but also to obtain a relevance ranking.

\subsection{Heatmap and Relevance Ranking}

Through BRF it is possible to obtain a ranking with the most relevant frequency bands in each analyzed \textit{k}-level. To facilitate the visualization of the frequency bands, a heatmap is proposed, where the values are normalized on a scale from 1 to -1, with only positive values being relevant.

Each level represents the division of the signal into {$2\textsuperscript{k}$}, where \textit{k} is the level and {$2\textsuperscript{k}$} is the amount of bands present in each level. Example: a complete signal with a frequency range from 0 to 10240 hz, for level 0, the bandwidth (BW) will be from 0 to 10240 hz ({$2\textsuperscript{0}$}), that is, the complete signal. For level 1, the signal will be divided into two bands ({$2\textsuperscript{1}$}), 0 to 5120 hz and 5120 to 10240 hz and so on for the other levels.

It is worth mentioning that the analysis must be performed per level, that is, each level will present a relevance ranking, since the calculations are performed separately, and only combined to obtain all the results in a single graph. Then the analysts can determine the size of the band (BW) on which to perform an analysis.

The same happens for the \textit{rms} value used for comparison, however the values are normalized from 0 to 1, as there are no negative \textit{rms} values in the frequency signal.

\section{Experimental procedure}
\subsection{Data description}

Three datasets were used: one synthetically created to exemplify the proposed approach (Case 1), one publicly available (Case 2 - Bearing dataset) and one developed by the author for the study (Case 3 - Mechanical faults dataset). The datasets were chosen because they contain different faults that are popular in rotating machinery, in addition to presenting the differences between normal and fault conditions for each rotating machinery. The use of different datasets allows to validate the proposed methodology under different conditions, and to show the strengths and limitations.

All real world datasets have vibration signals collected using accelerometers.

\subsubsection{Case 1: Synthetic dataset}

The proposed methodology aims to mainly evaluate the frequency bands that have relevance in the signal, thus allowing the analysis by the expert, or extraction of features for use in machine learning models. In this way, a synthetic dataset was generated, aiming to replicate real conditions in rotating machinery.

The proposed signal is an oscillatory signal whose waveform is given in the time domain by a function:

\begin{flalign}
    & x(t) = A\,sin(2\pi\,ft + \theta\,) & \label{eq:sinal_oscil_ref}
\end{flalign}

Where \textit{A} indicates the peak amplitude of the signal, \textit{f} the frequency [Hz], \textit{t} a time vector [s] and \textit{$\theta$} the phase [rad]; since monoaxial accelerometers without tachometer were used, the phase reference was not considered, although it could help in the identification of some defects. In our experiments \textit{t} = 1s and \textit{fs} = 20480 hz were used.

The harmonic signal was chosen to compose the dataset, due to its similarity with characteristic defects in rotating machinery, such as the force produced by the unbalance of a rotor. Furthermore, the harmonic signal represents the fundamental rotation frequency of the machine (1x), which is one of the most relevant parameter for the analysis of rotating machinery signals.

In real world signals, in addition to the influence of other frequencies related to nearby machines, external excitations etc., some vibrations/defects excite more than one frequency in the signal, such as mechanical looseness, misalignment, gear defects. To simulate a condition closer to reality, a synthetic signal was composed with \textit{n} different frequencies:

\begin{flalign}
    & x_t(t) = A_1\,sin(2\pi\,f_1t + \theta\,)+A_2\,sin(2\pi\,f_2t + \theta\,)+\ldots+A_n\,sin(2\pi\,f_nt + \theta\,) & \label{eq:sinal_oscil}
\end{flalign}

Where \textit{{$x\textsubscript{t}$}(t)} is a sum of harmonic signals \textit{x(t)} Eq.\eqref{eq:sinal_oscil_ref}, of different frequencies, {$f\textsubscript{1}$}, {$f\textsubscript{2}$},..., {$f\textsubscript{n}$}. The frequencies were defined in such a way that when performing the division by bands, some bands contained more than one frequency, others only one and others none, namely: 30, 120, 500, 700, 750, 2300, 2450, 2600, 2700, 2800 and 3450 hz. Amplitudes were randomly determined for a range of 0 - 1.

In real applications, every signal will present a noise level, and for this reason, it was added to the synthetic signal, as follows:

\begin{flalign}
    & x_f = x_t + \alpha \textsubscript{gauss} G, \, G \, \sim \, N(0,1) & \label{eq:gauss_noise}
\end{flalign}

where $\alpha \textsubscript{gauss} >$ 0 is the Gaussian noise coefficient, {$x\textsubscript{f}$} is the final synthetic signal, {$x\textsubscript{t}$} is the raw signal and \textit{G} the Gaussian noise.

As the proposed method aims at identifying relevant frequency bands, and at knowing that they can be influenced by the noise present in the signal, four noise levels were added.

As it is a synthetic and known signal, the different noise levels were obtained based on the signal-to-noise ratio (SNR). SNR is the ratio between the desired information or the power of a signal and the undesired signal or the power of the background noise, and its unit of expression is typically decibels (dB):

\begin{flalign}
    & SNR = 10{\log\textsubscript{10}} \left(\frac{P\textsubscript{signal}}{P\textsubscript{noise}}\right) & \label{eq:snr},
\end{flalign}
where, {P\textsubscript{signal}} and {P\textsubscript{noise}} are the \textit{rms} value in volts for the signal and for the noise respectively.

Observing Eq.\eqref{eq:snr}, it shoulde be noted that SNR greater than 1 (or greater than 0 dB) indicates more signal than noise. Therefore, the SNR levels were defined as: 24, 12, 6 and 0 dB. The SNR relationships were also designed in such a way to facilitate the visualization of the signals in the frequency domain, allowing for a better validation of the methodology, with the proposal being:

i) 24 dB (SNR): the harmonic frequencies present in the signal are easily evidenced in spectral analysis, with a small noise;

ii) 12 dB (SNR): harmonic frequencies continue to be visible in a spectral analysis, but with an increase in background noise, which does not preclude a correct identification;

iii) 6 dB (SNR): harmonic frequencies of greater amplitude are identifiable in a conventional spectral analysis, and may have some frequencies close to noise, which would raise doubts as to whether it is noise or harmonic frequency;

iv) 0 dB (SNR): harmonic frequencies and noise are completely mixed in a conventional spectral analysis, not allowing the identification of machine faults.

To facilitate the description in the text, the noises will be called \emph{low}, \emph{medium}, \emph{high} and \emph{mixed}, referring to SNR 24, 12, 6 and 0 dB, respectively.

\subsubsection{Case 2: Bearing dataset}

The dataset \cite{QIU20061066} is composed of remaining useful life (RUL) test on bearings, with 03 tests performed and 4 bearings in each test. The end of life reached for each bearing occurred after 100 million revolutions which is the designed life time. Each file consists of 20,480 points with the sampling rate set at 20 kHz. The signals were manually labeled, based on knowledge about fault diagnosis using vibration analysis, and 2 signals were selected to validate the applicability of the proposed method. For the study, the bearing 01 of test 02 was used, with signal 150 representing the normal operating condition and signal 905 with fault in the outer race.

Since the bearing dataset (Case 2) is publicly available and already explored in \cite{xai_lucas}, we refer the interest readers to \cite{QIU20061066, xai_lucas} for more details. 

\subsubsection{Case 3: Mechanical faults dataset}

\begin{figure*}[ht]
  \centering
  \includegraphics[scale=0.05]{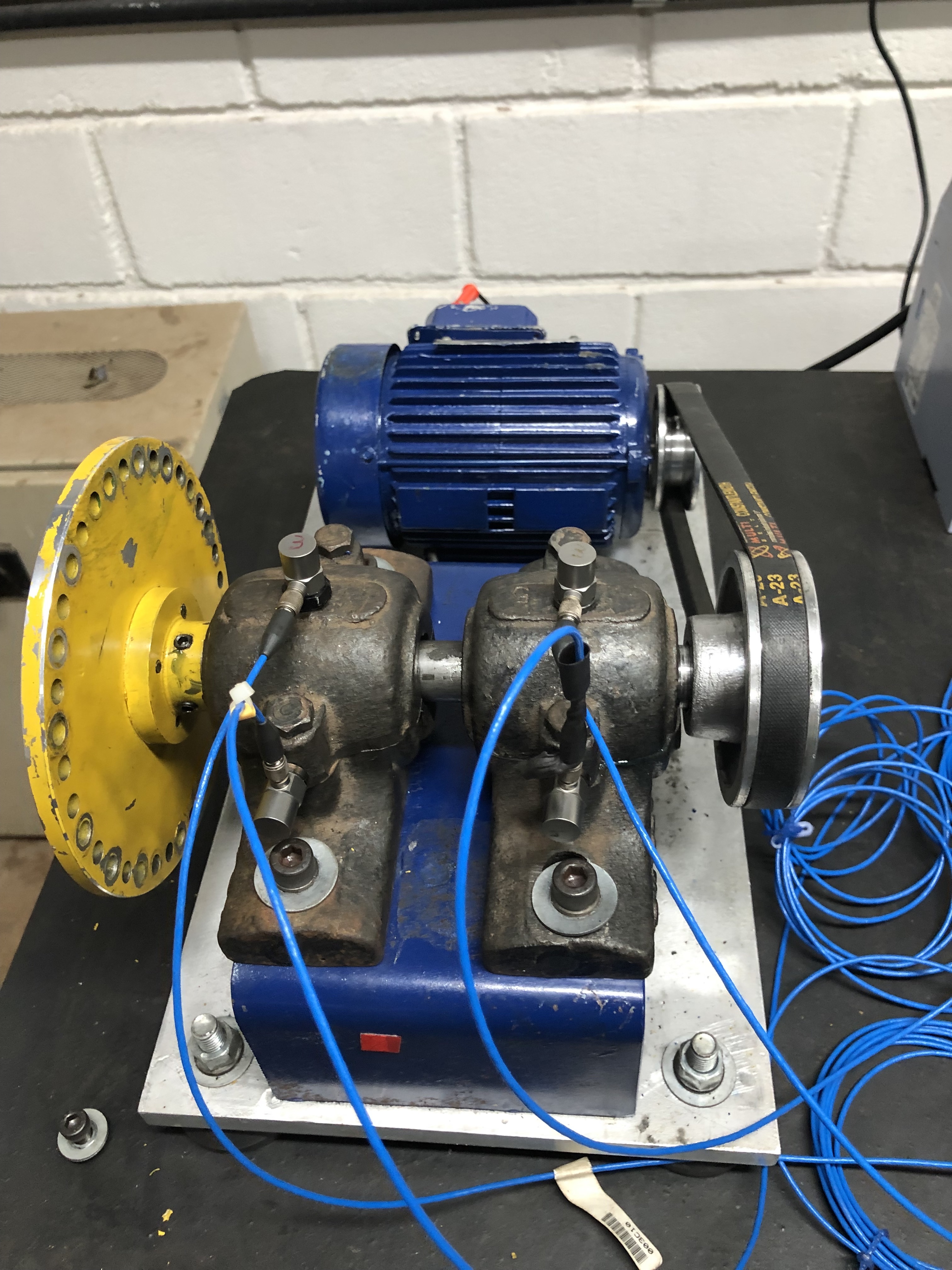}
  \caption{Bench test.}
  \label{fig:Bench}
\end{figure*}

The last dataset was developed by the authors. The faults were introduced on a test bench, depicted in Fig. \ref{fig:Bench} amd composed by: motor, frequency inverter, bearing house, two bearings, two pulleys, belt and rotor (disc).

20 tests were performed, 5 for each fault condition considered: unbalance, misalignment, mechanical looseness and normal operating condition. Each test consists of 4 sets of 420 signals collected continuously. Each file consisting of 25,000 points with the sampling rate set at 25 kHz (420 signals per accelerometer), resulting in a total of 8400 signals per accelerometer. The rotation was kept constant with a value measured on the axis of approximately 1238 rpm. The sequence of tests was randomly defined. Before starting any test, the bench was dismantled and returned to normal operating condition, to later introduce the fault. The experimental procedure allows variations to occur, making the tests closer to industrial reality.

Since the available bench can be considered small, the measurement points are close, and in order to reduce the computational cost of the tests, only the signals from the horizontal position of the accelerometer present in the coupled side bearing (near the pulley) were used in the analyses.

In addition, only two conditions were chosen to facilitate the reading and understanding of the methodology, in view of the other dataset used, namely: normal operating condition and unbalance.

\section{Results and discussion}
\subsection{Data exploration}

In this section, the synthetically generated samples and the two datasets used are analyzed and discussed. As the objective is just to visualize the signal characteristic, the signal in the frequency domain was plotted with \textit{df} = 1 hz.

\subsubsection{Case 1: Synthetic dataset}

\begin{figure}[ht]
  \subfloat[Low noise]{
	\begin{minipage}[c][0.75\width]{
	   0.23\textwidth}
	   \centering
       \label{fig:case1_low}
	   \includegraphics[width=1.15\textwidth]{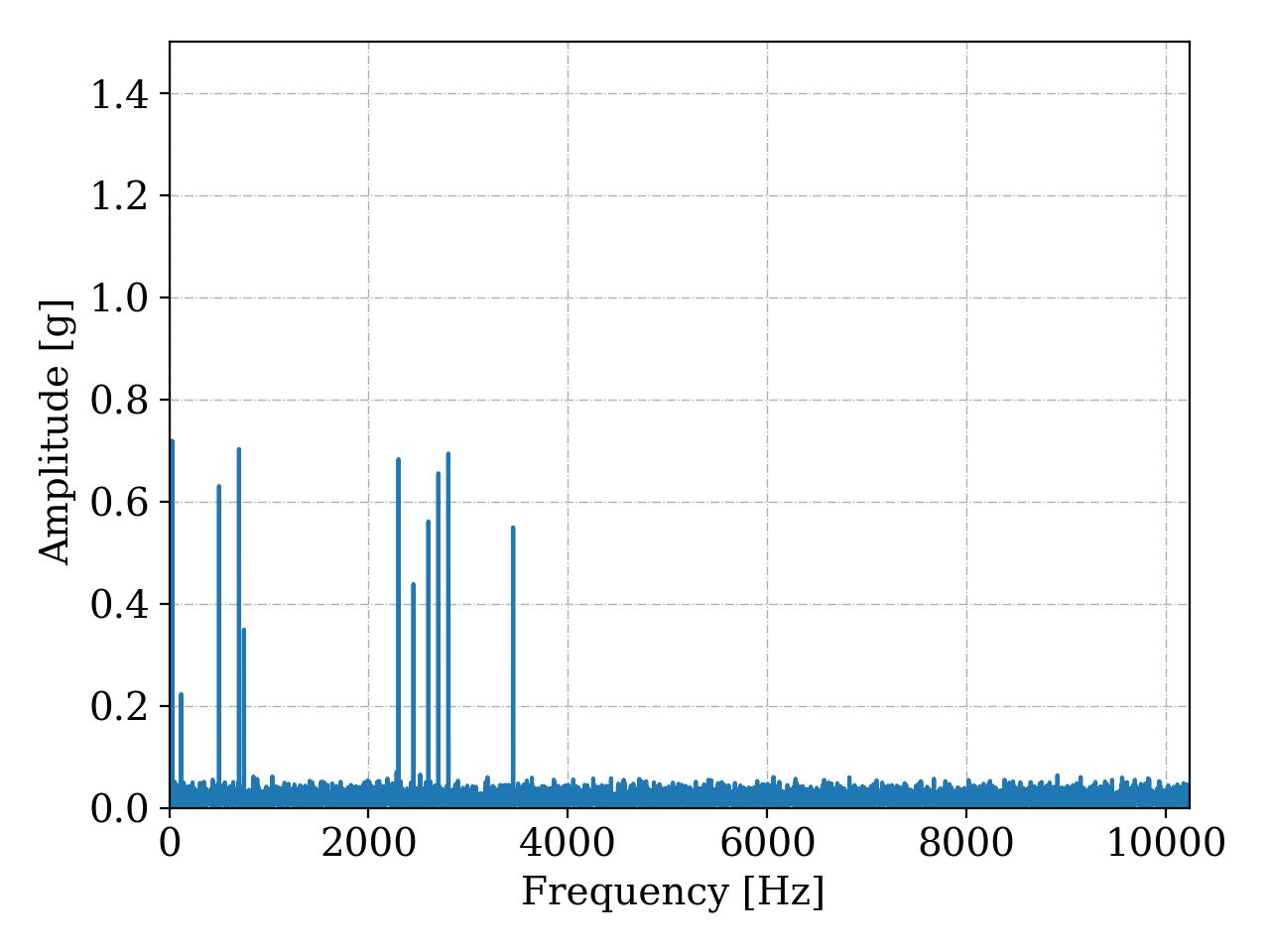}
	\end{minipage}}
 \hfill 
  \subfloat[Medium noise]{
	\begin{minipage}[c][0.75\width]{
	   0.23\textwidth}
	   \centering
        \label{fig:case1_med}
	   \includegraphics[width=1.15\textwidth]{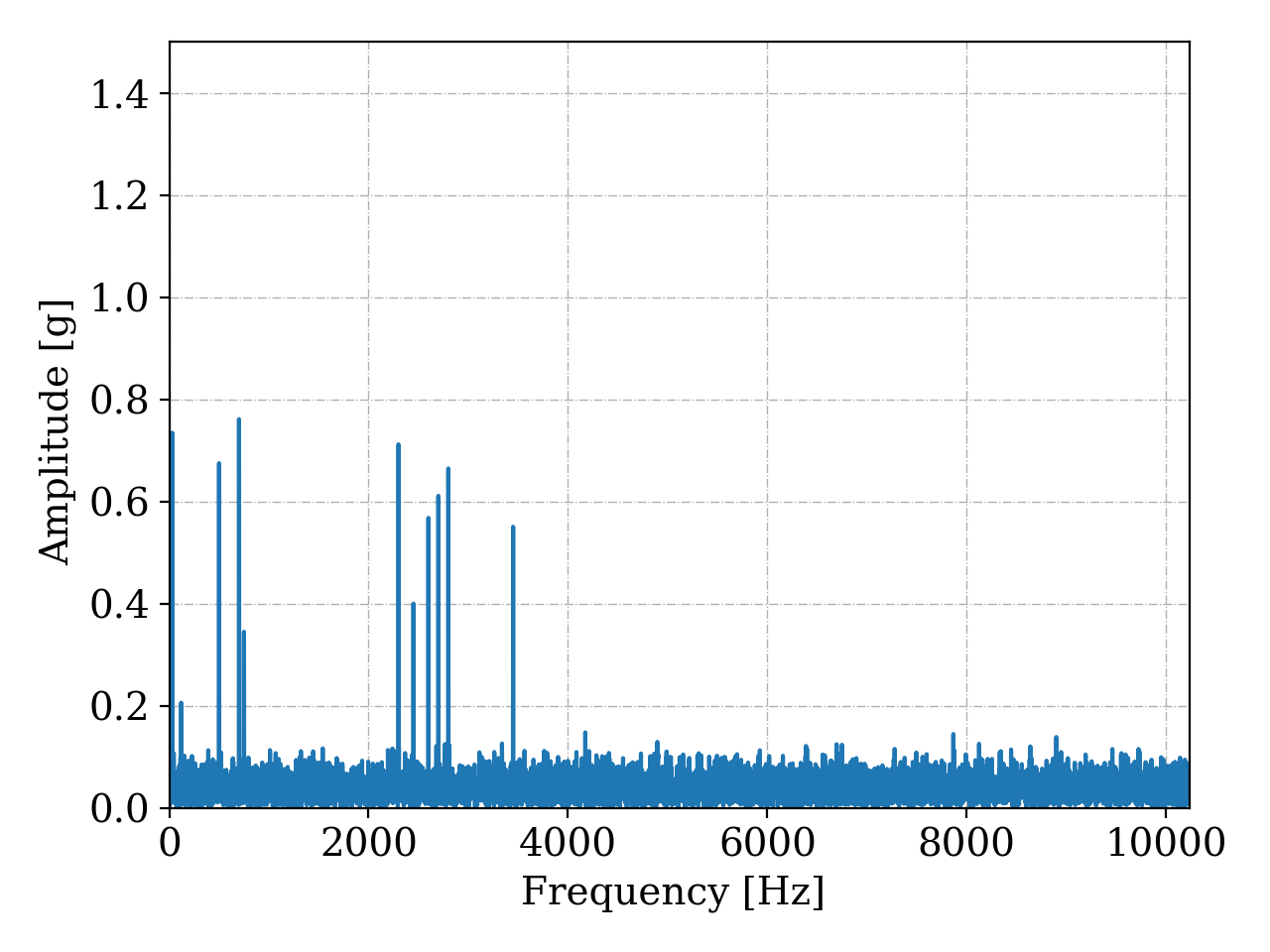}
	\end{minipage}} 
 \hfill 
  \subfloat[High noise]{
	\begin{minipage}[c][0.75\width]{
	   0.23\textwidth}
	   \centering
       \label{fig:case1_high}
	   \includegraphics[width=1.15\textwidth]{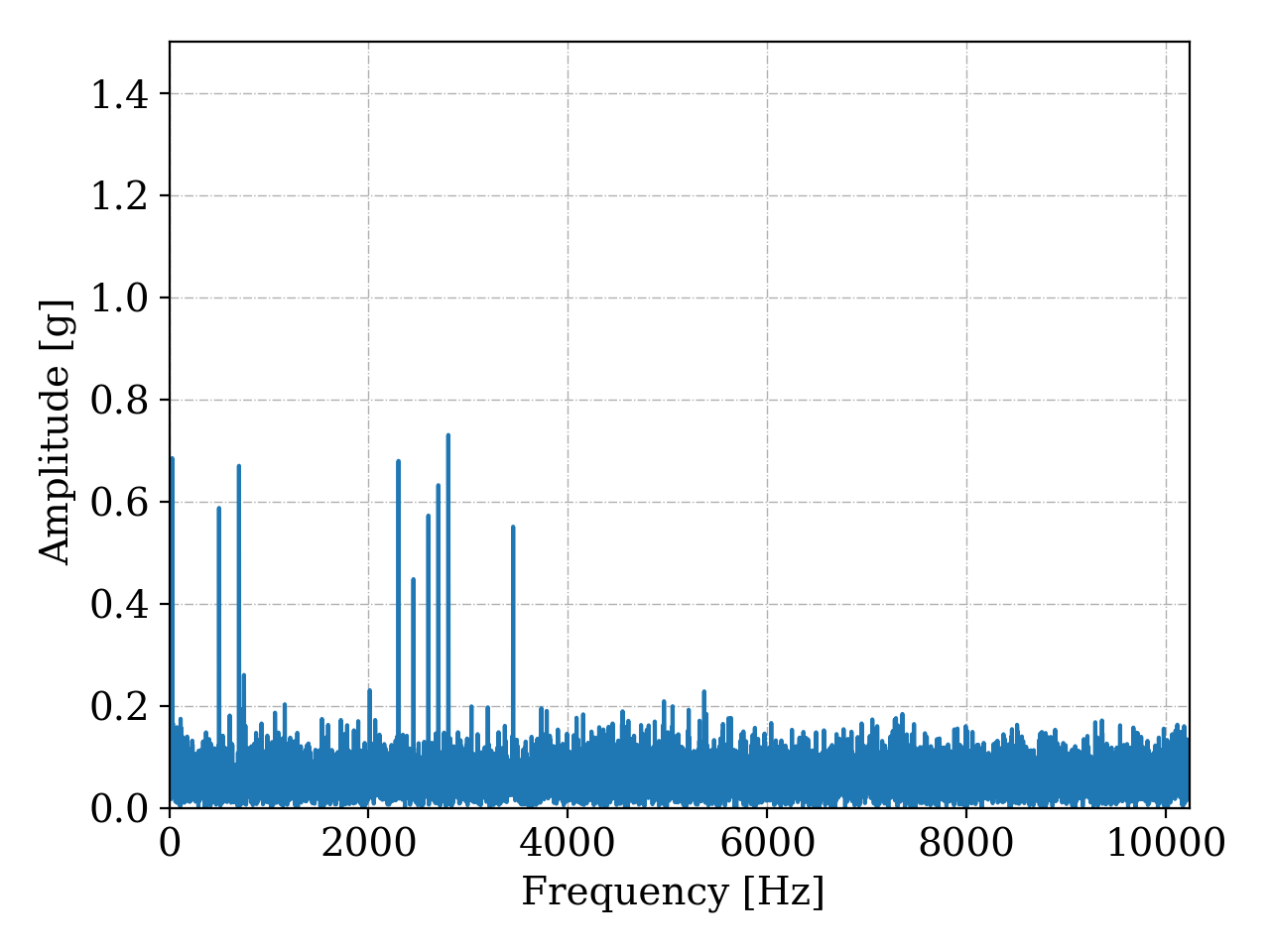}
	\end{minipage}}
 \hfill 
  \subfloat[Mixed noise]{
	\begin{minipage}[c][0.75\width]{
	   0.23\textwidth}
	   \centering
        \label{fig:case1_mix}
	   \includegraphics[width=1.15\textwidth]{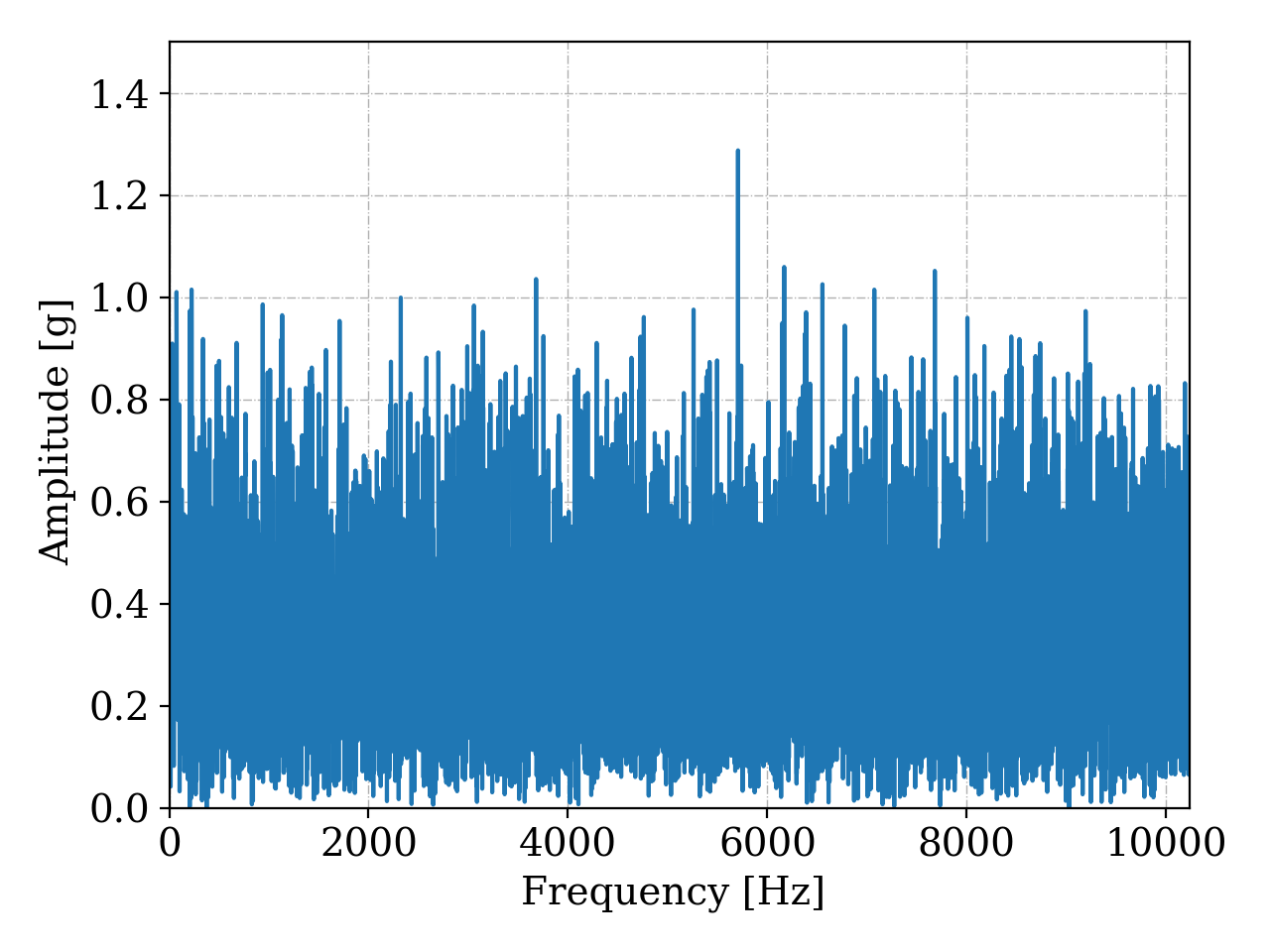}
	\end{minipage}} 
 \hfill 
    \caption{Examples of vibration signals for Case 1.}
    \label{fig:sinalcase1}	
\end{figure}

In Fig.\ref{fig:sinalcase1} the original signals for the two conditions (Normal and BPFO) and the synthetic signals for the seven created conditions (Normal, BPFO, BPFI, Unbalance, Misalignment, Looseness and Gear Fault) from Case 1 are presented.

\renewcommand{\baselinestretch}{3} 
\renewcommand{\baselinestretch}{1.5} 

In Fig.\ref{fig:case1_low},\ref{fig:case1_med},\ref{fig:case1_high} it can be verified that the generated frequencies are highlighted as expected, since the SNR relationship is positive, presenting more signal than noise in the sample. On the other hand, in Fig.\ref{fig:case1_mix} there is a mixture of the signal and noise not allowing to identify the frequencies of interest originally inserted.

In a conventional vibration analysis, focusing on the identification of faults (predictive maintenance), it can be concluded that Fig.\ref{fig:case1_low},\ref{fig:case1_med},\ref{fig:case1_high} allow a diagnosis of the signal, while Fig.\ref{fig:case1_mix} would be irrelevant since it is not possible to distinguish the characteristic fault frequencies from the noise. The same can be analyzed in relation to the frequency bands relevant to the diagnosis. The bands in which the frequencies are highlighted are usually related to the excitations in the machines and, consequently, are relevant for the diagnosis of the defect.

In a visual vibration analysis, stand out the bands containing the frequencies: 30, 120, 500, 700, 750, 2300, 2450, 2600, 2700, 2800 and 3450 Hz (which were synthetically inserted in the signal). Other regions in the signals represent only noise. Thus, the methodology must be able to identify the bands in which such frequencies are present.

\subsubsection{Case 2: Bearing Fault}

\begin{figure}[ht]
  \subfloat[Normal condition]{
	\begin{minipage}[c][0.35\width]{
	   0.5\textwidth}
	   \centering
       \label{fig:case2_normal}
	   \includegraphics[width=0.55\textwidth]{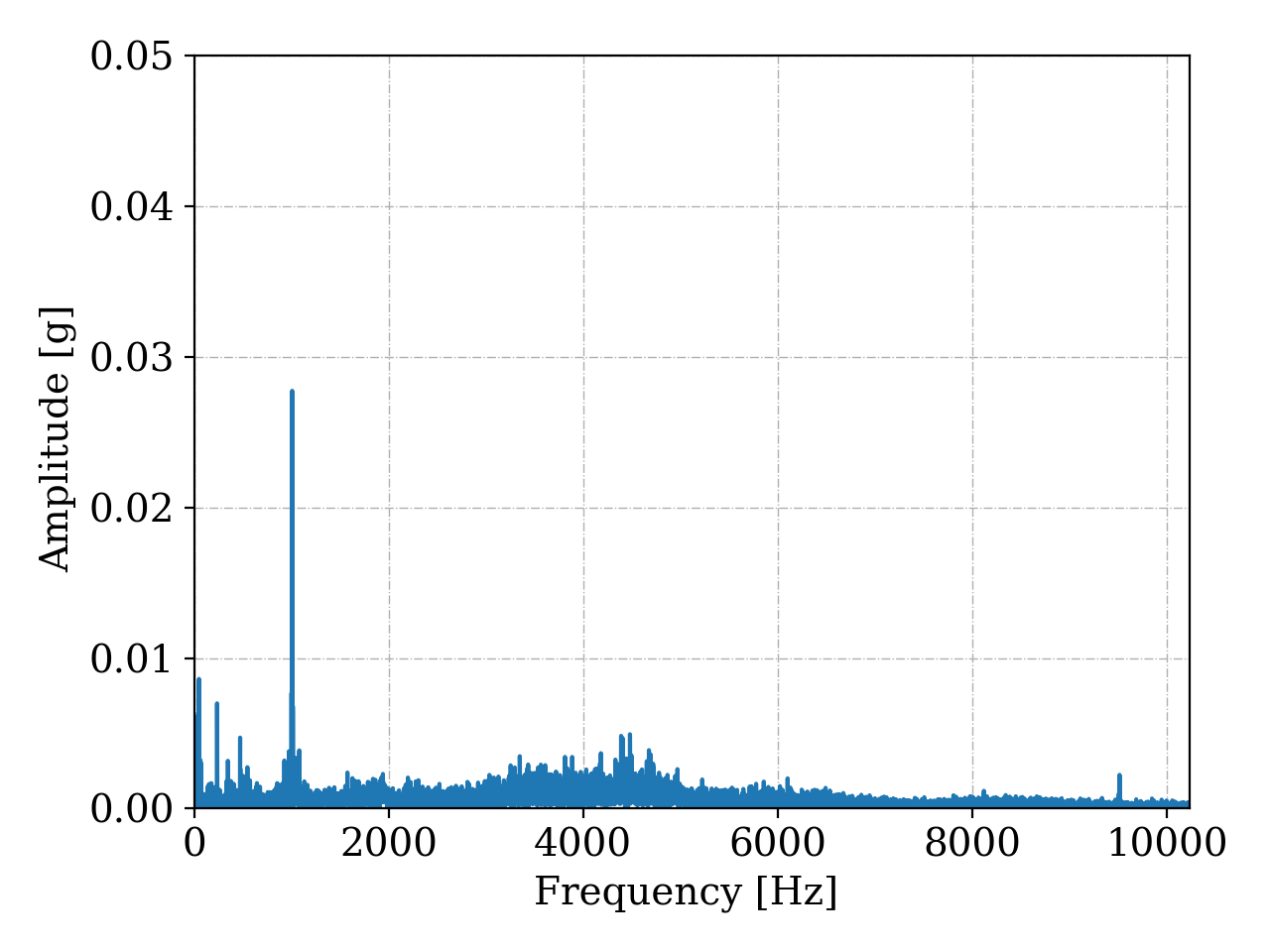}
	\end{minipage}}
 \hfill 
  \subfloat[Outer race fault]{
	\begin{minipage}[c][0.35\width]{
	   0.5\textwidth}
	   \centering
        \label{fig:case2_fault}
	   \includegraphics[width=0.55\textwidth]{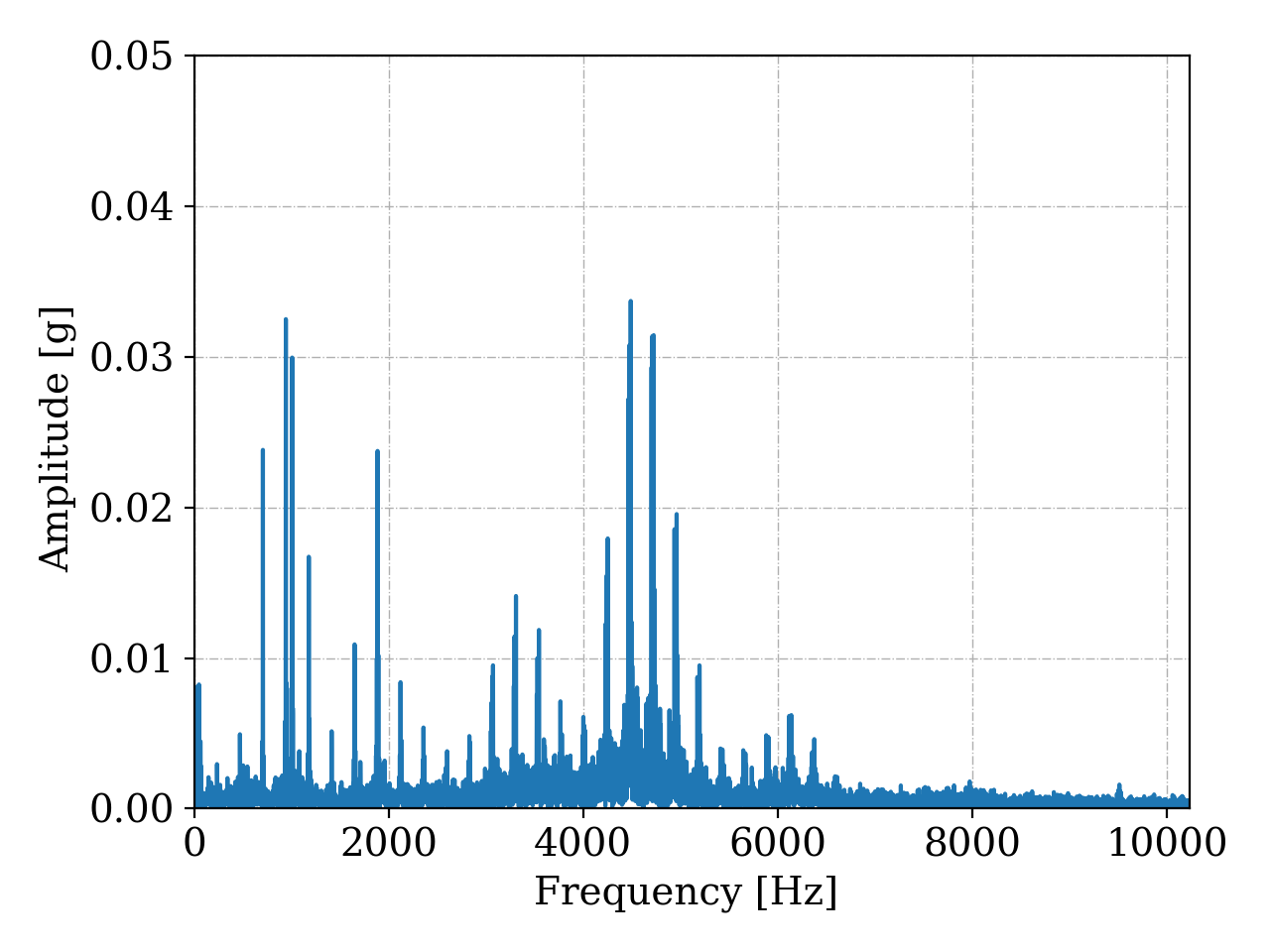}
	\end{minipage}} 
 \hfill 
    \vspace*{-7mm}
    \caption{Examples of vibration signals for Case 2.}
    \label{fig:sinalcase2}	
\end{figure}

In Fig.\ref{fig:sinalcase2}, the two randomly selected signals from the Case 2, representing normal operating condition Fig.\ref{fig:case2_normal} and bearing outer race fault Fig.\ref{fig:case2_fault} are presented.

\renewcommand{\baselinestretch}{3} 
\renewcommand{\baselinestretch}{1.5} 

In Fig.\ref{fig:case2_normal} it can be noted that only one frequency related to the dynamic behavior of the equipment is highlighted since there is no fault in the signal. On the other hand, in Fig.\ref{fig:case2_fault} other frequencies become present, indicating to the specialist a change in equipment behavior, which in turn is related to the fault. The frequencies highlighted in both cases are relevant for diagnosis and should be analyzed by the specialist to determine the current condition of the equipment. Therefore, as in the analysis, the methodology must be able to identify such frequencies as relevant.

\subsubsection{Case 3: Mechanical faults dataset}

\begin{figure}[ht]
  \subfloat[Normal condition]{
	\begin{minipage}[c][0.35\width]{
	   0.5\textwidth}
	   \centering
       \label{fig:case3_normal}
	   \includegraphics[width=0.55\textwidth]{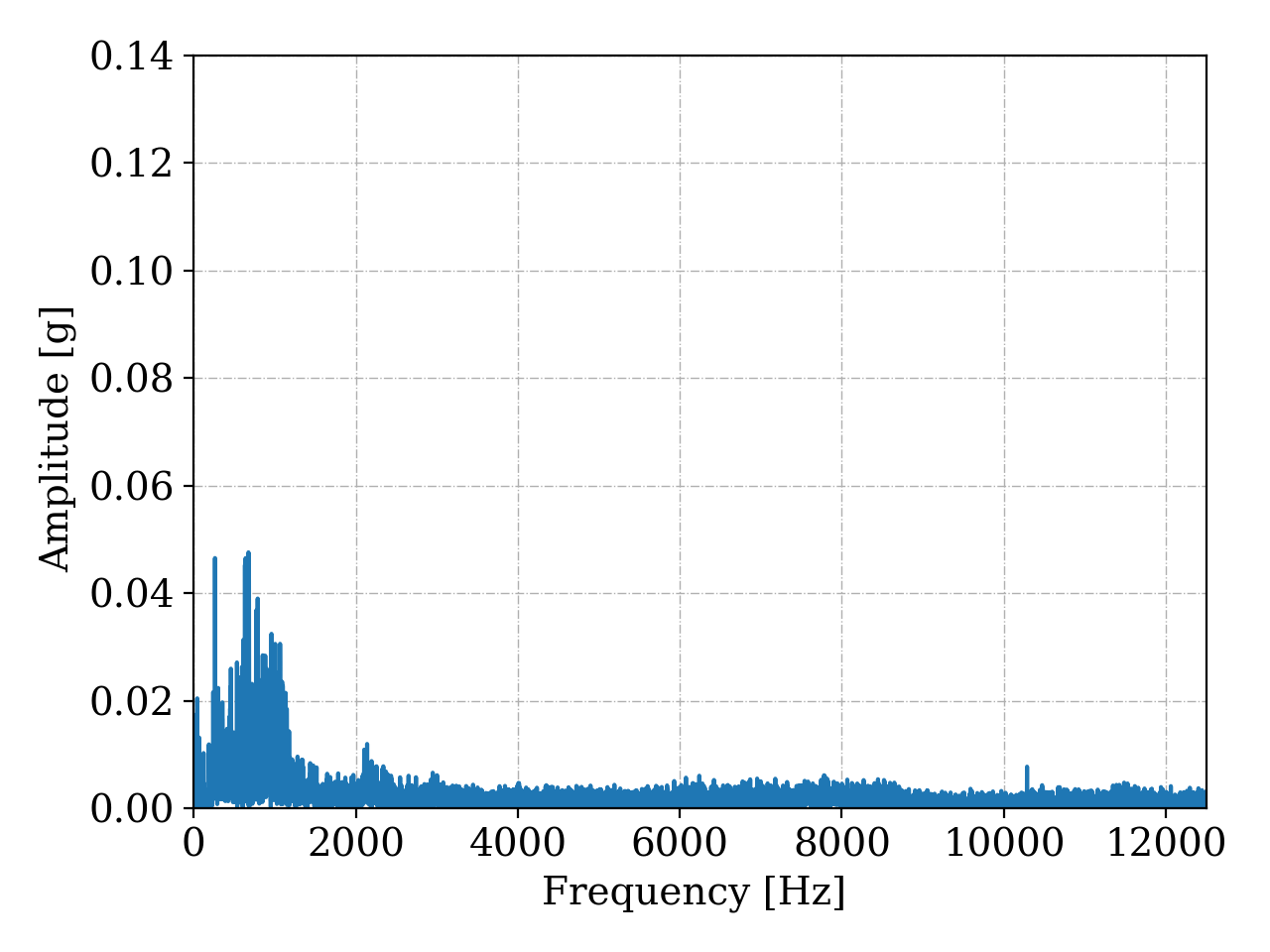}
	\end{minipage}}
 \hfill 
  \subfloat[Unbalance]{
	\begin{minipage}[c][0.35\width]{
	   0.5\textwidth}
	   \centering
        \label{fig:case3_fault}
	   \includegraphics[width=0.55\textwidth]{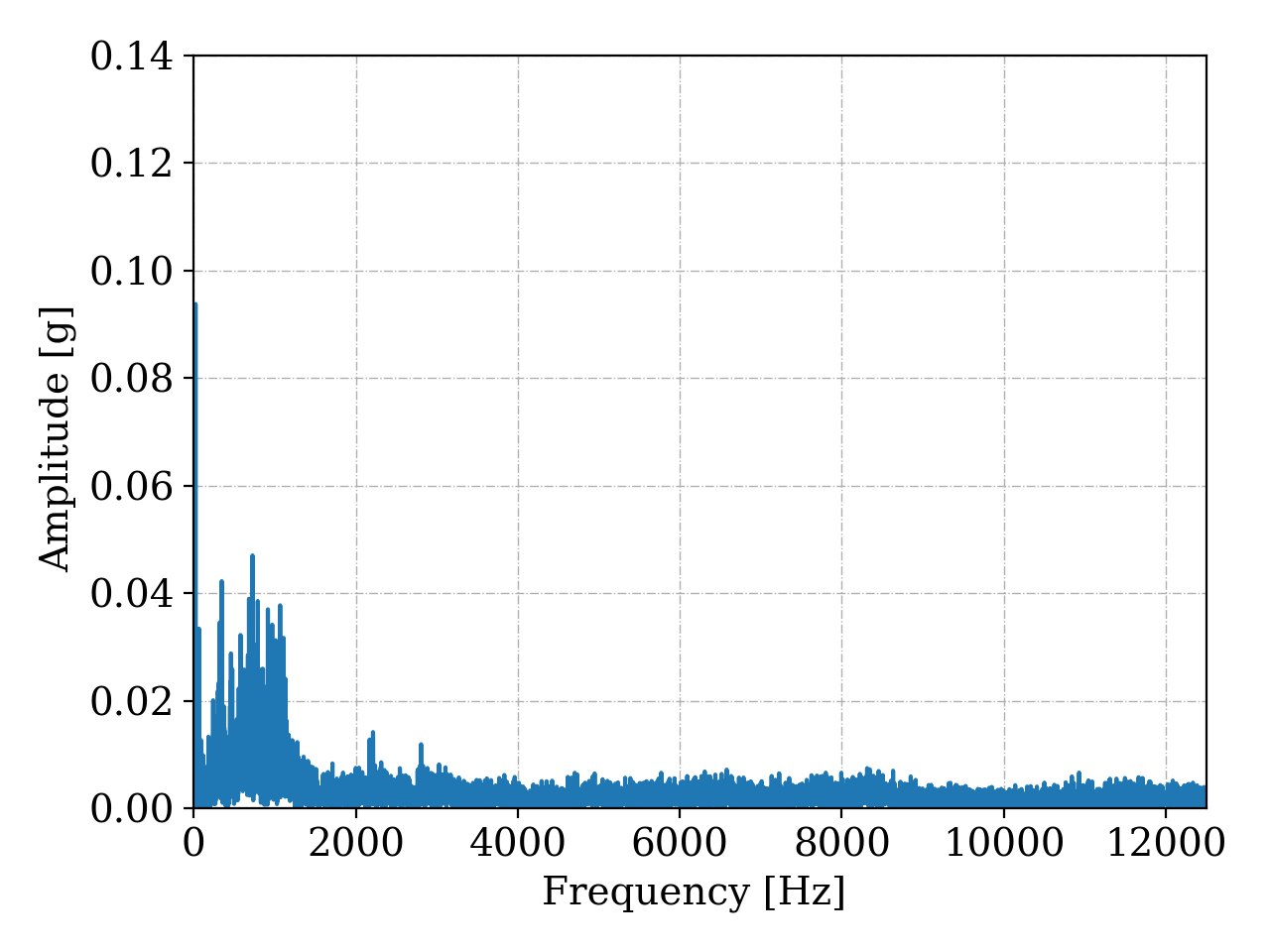}
	\end{minipage}} 
 \hfill 
    \vspace*{-7mm}
    \caption{Examples of vibration signals for Case 3.}
    \label{fig:sinalcase3}	
\end{figure}

In Fig.\ref{fig:sinalcase3}, the two randomly selected signals from the Case 3, representing normal operating condition Fig.\ref{fig:case3_normal} and unbalance Fig.\ref{fig:case3_fault} are presented.

\renewcommand{\baselinestretch}{3} 
\renewcommand{\baselinestretch}{1.5} 

The dataset was purposely generated to simulate severe industrial conditions, where the vibration signal has a rich amount of frequencies, even if they are not related to equipment faults (e.g., instrumentation interference, nearby equipment etc.). Thus, even for the normal operating condition, it can be noted that in Fig.\ref{fig:case3_normal} there are several highlighted frequencies. In Fig.\ref{fig:case3_fault} where the unbalance was inserted, it is verified that the frequency related to the unbalance starts to dominate the signal, presenting greater relevance for the analysis (21.9 hz). Through a vibration analysis performed by a specialist, the rotation frequency will be related to the characteristic of the unbalance defect in the machine. Therefore, the methodology must be able to identify such frequency as relevant even with the presence of other frequencies in the signal.

\subsection{BRF Analysis}

In this subsection, the results obtained with the proposed methodology and \textit{rms} value are presented. Due to the lack of space, for each dataset a signal was selected and the heatmap with the most relevant bands per level is presented. For the analysis of the other signals, the ranking with the five most relevant bands are presented.

\subsubsection{Case 1: Synthetic dataset}

\begin{figure*}[ht]
  \centering
  \includegraphics[scale=0.55]{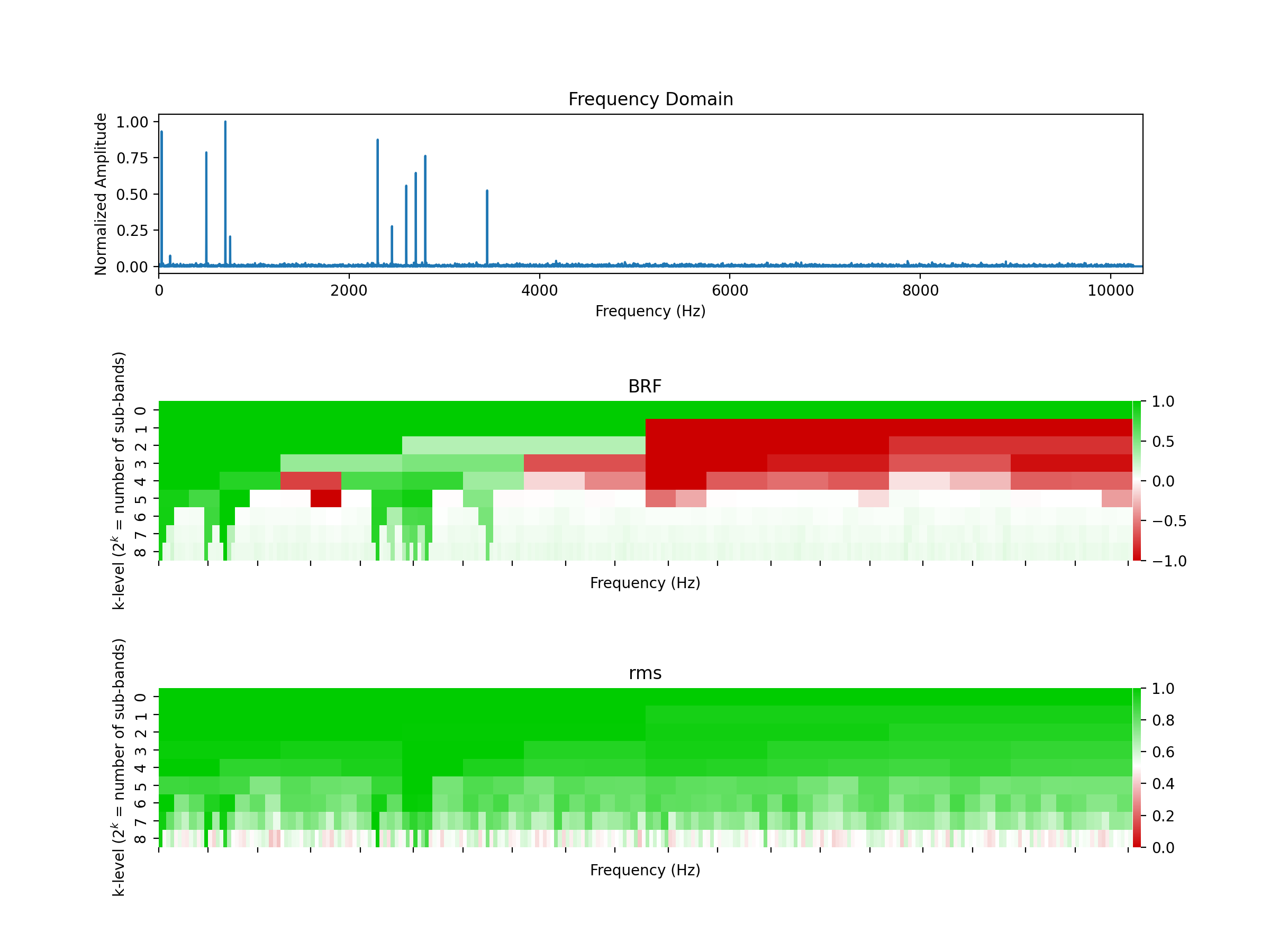}
  \caption{Heatmap - Case 1.}
  \label{fig:case1_heatmap}
\end{figure*}

To represent the synthetic dataset, the heatmap with medium noise was selected, Fig. \ref{fig:case1_heatmap}.

It can be noted that the BRF was able to identify as relevant the bands that contain the frequencies inserted in the signal. It is also verified that the smaller the size of the band, the BRF tends to classify regions that present only noise as relevant. This occurs because, as it is limited in a small region, only one or a few dominant frequencies can occur and the entropy may assumes a value close to a system with equiprobable signal. This limitation is circumvented by analyzing the ranking with the most relevant bands determined by the method (or heatmap color scale), which points out to the harmonic frequencies inserted in the signal.

Comparing with the \textit{rms}, the methodology was able to better distinguish the relevant bands for analysis, as it can be observed in levels 1, 2, 3, 4 and 5, mainly. According to the BRF, bands with a value less than/equal to zero (scale from white to red) are irrelevant for the analysis. Analyzing the heatmap, the regions with a white to red scale are bands where there is only the presence of noise and no harmonic frequency. On the other hand, the same bands are classified with maximum value or close to maximum by the \textit{rms} method.

The rankings with the five most relevant bands according to BRF and \textit{rms} for the four signals from Case 1 are shown in Table~\ref{tab:ranking_case1}.

\begin{table}[]
\caption{}
\caption*{Relevance Ranking - Case 1}
\label{tab:ranking_case1}
\resizebox{\textwidth}{!}{
\begin{tabular}{@{}cc|ccccccccc@{}}
\hline
\textbf{Signal}                          & \textbf{Feat.}                                         & \textbf{\begin{tabular}[c]{@{}c@{}}Rank/\\ k-level\end{tabular}} & \textbf{\begin{tabular}[c]{@{}c@{}}1\\ (BW 5120 hz)\end{tabular}} & \textbf{\begin{tabular}[c]{@{}c@{}}2\\ (BW 2560 hz)\end{tabular}} & \textbf{\begin{tabular}[c]{@{}c@{}}3\\ (BW 1280 hz)\end{tabular}} & \textbf{\begin{tabular}[c]{@{}c@{}}4\\ (BW 640hz)\end{tabular}} & \textbf{\begin{tabular}[c]{@{}c@{}}5\\ (BW 320 hz)\end{tabular}} & \textbf{\begin{tabular}[c]{@{}c@{}}6\\ (BW 160 hz)\end{tabular}} & \textbf{\begin{tabular}[c]{@{}c@{}}7\\ (BW 80 hz)\end{tabular}} & \textbf{\begin{tabular}[c]{@{}c@{}}8\\ (BW 40 hz)\end{tabular}} \\ \hline
                                         & \cellcolor[HTML]{F2F2F2}                               & \cellcolor[HTML]{F2F2F2}\textbf{1}                               & \cellcolor[HTML]{F2F2F2}0:5120                                    & \cellcolor[HTML]{F2F2F2}0:2560                                    & \cellcolor[HTML]{F2F2F2}0:1280                                    & \cellcolor[HTML]{F2F2F2}2560:3200                               & \cellcolor[HTML]{F2F2F2}2560:2880                                & \cellcolor[HTML]{F2F2F2}0:160                                    & \cellcolor[HTML]{F2F2F2}0:80                                    & \cellcolor[HTML]{F2F2F2}0:40                                    \\
                                         & \cellcolor[HTML]{F2F2F2}                               & \textbf{2}                                                       & -                                                                 & 2560:5120                                                         & 2560:3840                                                         & 0:640                                                           & 0:320                                                            & 2720:2880                                                        & 2800:2880                                                       & 2800:2840                                                       \\
                                         & \cellcolor[HTML]{F2F2F2}                               & \cellcolor[HTML]{F2F2F2}\textbf{3}                               & \cellcolor[HTML]{F2F2F2}-                                         & \cellcolor[HTML]{F2F2F2}-                                         & \cellcolor[HTML]{F2F2F2}1280:2560                                 & \cellcolor[HTML]{F2F2F2}640:1280                                & \cellcolor[HTML]{F2F2F2}640:960                                  & \cellcolor[HTML]{F2F2F2}640:800                                  & \cellcolor[HTML]{F2F2F2}640:720                                 & \cellcolor[HTML]{F2F2F2}680:720                                 \\
                                         & \cellcolor[HTML]{F2F2F2}                               & \textbf{4}                                                       & -                                                                 & -                                                                 & -                                                                 & 1920:2560                                                       & 2240:2560                                                        & 2240:2400                                                        & 2240:2320                                                       & 2280:2320                                                       \\
                                         & \multirow{-5}{*}{\cellcolor[HTML]{F2F2F2}\textbf{BRF}} & \cellcolor[HTML]{F2F2F2}\textbf{5}                               & \cellcolor[HTML]{F2F2F2}-                                         & \cellcolor[HTML]{F2F2F2}-                                         & \cellcolor[HTML]{F2F2F2}-                                         & \cellcolor[HTML]{F2F2F2}3200:3840                               & \cellcolor[HTML]{F2F2F2}320:640                                  & \cellcolor[HTML]{F2F2F2}2560:2720                                & \cellcolor[HTML]{F2F2F2}2640:2720                               & \cellcolor[HTML]{F2F2F2}2680:2720                               \\
                                         & \cellcolor[HTML]{D9E2F3}                               & \cellcolor[HTML]{D9E2F3}\textbf{1}                               & \cellcolor[HTML]{D9E2F3}0:5120                                    & \cellcolor[HTML]{D9E2F3}0:2560                                    & \cellcolor[HTML]{D9E2F3}0:1280                                    & \cellcolor[HTML]{D9E2F3}0:640                                   & \cellcolor[HTML]{D9E2F3}2560:2880                                & \cellcolor[HTML]{D9E2F3}2560:2720                                & \cellcolor[HTML]{D9E2F3}640:720                                 & \cellcolor[HTML]{D9E2F3}680:720                                 \\
                                         & \cellcolor[HTML]{D9E2F3}                               & \textbf{2}                                                       & 5120:10240                                                        & 2560:5120                                                         & 2560:3840                                                         & 2560:3200                                                       & 2240:2560                                                        & 640:800                                                          & 0:80                                                            & 0:40                                                            \\
                                         & \cellcolor[HTML]{D9E2F3}                               & \cellcolor[HTML]{D9E2F3}\textbf{3}                               & \cellcolor[HTML]{D9E2F3}-                                         & \cellcolor[HTML]{D9E2F3}5120:7680                                 & \cellcolor[HTML]{D9E2F3}1280:2560                                 & \cellcolor[HTML]{D9E2F3}1920:2560                               & \cellcolor[HTML]{D9E2F3}640: 960                                 & \cellcolor[HTML]{D9E2F3}0:160                                    & \cellcolor[HTML]{D9E2F3}2640:2720                               & \cellcolor[HTML]{D9E2F3}2680:2720                               \\
                                         & \cellcolor[HTML]{D9E2F3}                               & \textbf{4}                                                       & -                                                                 & 7680:10240                                                        & 8960:10240                                                        & 640:1280                                                        & 0:320                                                            & 480:640                                                          & 480:560                                                         & 2280: 2320                                                      \\
\multirow{-10}{*}{\textbf{Low Noise}}    & \multirow{-5}{*}{\cellcolor[HTML]{D9E2F3}\textbf{RMS}} & \cellcolor[HTML]{D9E2F3}\textbf{5}                               & \cellcolor[HTML]{D9E2F3}-                                         & \cellcolor[HTML]{D9E2F3}-                                         & \cellcolor[HTML]{D9E2F3}5120:6400                                 & \cellcolor[HTML]{D9E2F3}9600:10240                              & \cellcolor[HTML]{D9E2F3}320:640                                  & \cellcolor[HTML]{D9E2F3}2240:2400                                & \cellcolor[HTML]{D9E2F3}2240: 2320                              & \cellcolor[HTML]{D9E2F3}480:520                                 \\ \hline
                                         & \cellcolor[HTML]{F2F2F2}                               & \cellcolor[HTML]{F2F2F2}\textbf{1}                               & \cellcolor[HTML]{F2F2F2}0:5120                                    & \cellcolor[HTML]{F2F2F2}0:2560                                    & \cellcolor[HTML]{F2F2F2}0:1280                                    & \cellcolor[HTML]{F2F2F2}0:640                                   & \cellcolor[HTML]{F2F2F2}640:960                                  & \cellcolor[HTML]{F2F2F2}640:800                                  & \cellcolor[HTML]{F2F2F2}640:720                                 & \cellcolor[HTML]{F2F2F2}680:720                                 \\
                                         & \cellcolor[HTML]{F2F2F2}                               & \textbf{2}                                                       & -                                                                 & 2560:5120                                                         & 2560:3840                                                         & 640:1280                                                        & 2560:2880                                                        & 0:160                                                            & 0:80                                                            & 0:40                                                            \\
                                         & \cellcolor[HTML]{F2F2F2}                               & \cellcolor[HTML]{F2F2F2}\textbf{3}                               & \cellcolor[HTML]{F2F2F2}-                                         & \cellcolor[HTML]{F2F2F2}-                                         & \cellcolor[HTML]{F2F2F2}1280:2560                                 & \cellcolor[HTML]{F2F2F2}2560:3200                               & \cellcolor[HTML]{F2F2F2}0:320                                    & \cellcolor[HTML]{F2F2F2}2240:2400                                & \cellcolor[HTML]{F2F2F2}2240:2320                               & \cellcolor[HTML]{F2F2F2}2280:2320                               \\
                                         & \cellcolor[HTML]{F2F2F2}                               & \textbf{4}                                                       & -                                                                 & -                                                                 & -                                                                 & 1920:2560                                                       & 2240:2560                                                        & 480:640                                                          & 480:560                                                         & 480:520                                                         \\
                                         & \multirow{-5}{*}{\cellcolor[HTML]{F2F2F2}\textbf{BRF}} & \cellcolor[HTML]{F2F2F2}\textbf{5}                               & \cellcolor[HTML]{F2F2F2}-                                         & \cellcolor[HTML]{F2F2F2}-                                         & \cellcolor[HTML]{F2F2F2}-                                         & \cellcolor[HTML]{F2F2F2}3200:3840                               & \cellcolor[HTML]{F2F2F2}320:640                                  & \cellcolor[HTML]{F2F2F2}2720:2880                                & \cellcolor[HTML]{F2F2F2}2800:2880                               & \cellcolor[HTML]{F2F2F2}2800:2840                               \\
                                         & \cellcolor[HTML]{D9E2F3}                               & \cellcolor[HTML]{D9E2F3}\textbf{1}                               & \cellcolor[HTML]{D9E2F3}0:5120                                    & \cellcolor[HTML]{D9E2F3}0:2560                                    & \cellcolor[HTML]{D9E2F3}2560:3840                                 & \cellcolor[HTML]{D9E2F3}0:640                                   & \cellcolor[HTML]{D9E2F3}2560:2880                                & \cellcolor[HTML]{D9E2F3}0:160                                    & \cellcolor[HTML]{D9E2F3}2240: 2320                              & \cellcolor[HTML]{D9E2F3}480:520                                 \\
                                         & \cellcolor[HTML]{D9E2F3}                               & \textbf{2}                                                       & 5120:10240                                                        & 2560:5120                                                         & 0:1280                                                            & 2560:3200                                                       & 2240:2560                                                        & 2560:2720                                                        & 480:560                                                         & 2280:2320                                                       \\
                                         & \cellcolor[HTML]{D9E2F3}                               & \cellcolor[HTML]{D9E2F3}\textbf{3}                               & \cellcolor[HTML]{D9E2F3}-                                         & \cellcolor[HTML]{D9E2F3}5120:7680                                 & \cellcolor[HTML]{D9E2F3}1280:2560                                 & \cellcolor[HTML]{D9E2F3}1920:2560                               & \cellcolor[HTML]{D9E2F3}320:640                                  & \cellcolor[HTML]{D9E2F3}2720:2880                                & \cellcolor[HTML]{D9E2F3}640:720                                 & \cellcolor[HTML]{D9E2F3}0:40                                    \\
                                         & \cellcolor[HTML]{D9E2F3}                               & \textbf{4}                                                       & -                                                                 & 7680:10240                                                        & 5120:6400                                                         & 3200:3840                                                       & 0:320                                                            & 640:800                                                          & 0:80                                                            & 680:720                                                         \\
\multirow{-10}{*}{\textbf{Medium Noise}} & \multirow{-5}{*}{\cellcolor[HTML]{D9E2F3}\textbf{RMS}} & \cellcolor[HTML]{D9E2F3}\textbf{5}                               & \cellcolor[HTML]{D9E2F3}-                                         & \cellcolor[HTML]{D9E2F3}-                                         & \cellcolor[HTML]{D9E2F3}3840:5120                                 & \cellcolor[HTML]{D9E2F3}5120:5760                               & \cellcolor[HTML]{D9E2F3}640: 960                                 & \cellcolor[HTML]{D9E2F3}2240:2400                                & \cellcolor[HTML]{D9E2F3}2640:2720                               & \cellcolor[HTML]{D9E2F3}2680:2720                               \\ \hline
                                         & \cellcolor[HTML]{F2F2F2}                               & \cellcolor[HTML]{F2F2F2}\textbf{1}                               & \cellcolor[HTML]{F2F2F2}0:5120                                    & \cellcolor[HTML]{F2F2F2}0:2560                                    & \cellcolor[HTML]{F2F2F2}2560:3840                                 & \cellcolor[HTML]{F2F2F2}2560:3200                               & \cellcolor[HTML]{F2F2F2}2560:2880                                & \cellcolor[HTML]{F2F2F2}2720:2880                                & \cellcolor[HTML]{F2F2F2}2800:2880                               & \cellcolor[HTML]{F2F2F2}2800:2840                               \\
                                         & \cellcolor[HTML]{F2F2F2}                               & \textbf{2}                                                       & -                                                                 & 2560:5120                                                         & 0:1280                                                            & 0:640                                                           & 2240:2560                                                        & 2560:2720                                                        & 0:80                                                            & 0:40                                                            \\
                                         & \cellcolor[HTML]{F2F2F2}                               & \cellcolor[HTML]{F2F2F2}\textbf{3}                               & \cellcolor[HTML]{F2F2F2}-                                         & \cellcolor[HTML]{F2F2F2}-                                         & \cellcolor[HTML]{F2F2F2}1280:2560                                 & \cellcolor[HTML]{F2F2F2}1920:2560                               & \cellcolor[HTML]{F2F2F2}0:320                                    & \cellcolor[HTML]{F2F2F2}0:160                                    & \cellcolor[HTML]{F2F2F2}2240: 2320                              & \cellcolor[HTML]{F2F2F2}2280:2320                               \\
                                         & \cellcolor[HTML]{F2F2F2}                               & \textbf{4}                                                       & -                                                                 & -                                                                 & -                                                                 & 640:1280                                                        & 640: 960                                                         & 2240:2400                                                        & 640:720                                                         & 680:720                                                         \\
                                         & \multirow{-5}{*}{\cellcolor[HTML]{F2F2F2}\textbf{BRF}} & \cellcolor[HTML]{F2F2F2}\textbf{5}                               & \cellcolor[HTML]{F2F2F2}-                                         & \cellcolor[HTML]{F2F2F2}-                                         & \cellcolor[HTML]{F2F2F2}-                                         & \cellcolor[HTML]{F2F2F2}3200:3840                               & \cellcolor[HTML]{F2F2F2}320:640                                  & \cellcolor[HTML]{F2F2F2}640:800                                  & \cellcolor[HTML]{F2F2F2}2640:2720                               & \cellcolor[HTML]{F2F2F2}2680:2720                               \\
                                         & \cellcolor[HTML]{D9E2F3}                               & \cellcolor[HTML]{D9E2F3}\textbf{1}                               & \cellcolor[HTML]{D9E2F3}0:5120                                    & \cellcolor[HTML]{D9E2F3}2560:5120                                 & \cellcolor[HTML]{D9E2F3}2560:3840                                 & \cellcolor[HTML]{D9E2F3}0:640                                   & \cellcolor[HTML]{D9E2F3}2560:2880                                & \cellcolor[HTML]{D9E2F3}2560:2720                                & \cellcolor[HTML]{D9E2F3}480:560                                 & \cellcolor[HTML]{D9E2F3}2280:2320                               \\
                                         & \cellcolor[HTML]{D9E2F3}                               & \textbf{2}                                                       & 5120:10240                                                        & 0:2560                                                            & 0:1280                                                            & 2560:3200                                                       & 4160:4480                                                        & 640:800                                                          & 2560:2640                                                       & 480:520                                                         \\
                                         & \cellcolor[HTML]{D9E2F3}                               & \cellcolor[HTML]{D9E2F3}\textbf{3}                               & \cellcolor[HTML]{D9E2F3}-                                         & \cellcolor[HTML]{D9E2F3}5120:7680                                 & \cellcolor[HTML]{D9E2F3}3840:5120                                 & \cellcolor[HTML]{D9E2F3}5120:5760                               & \cellcolor[HTML]{D9E2F3}0:320                                    & \cellcolor[HTML]{D9E2F3}3680:3840                                & \cellcolor[HTML]{D9E2F3}0:80                                    & \cellcolor[HTML]{D9E2F3}0:40                                    \\
                                         & \cellcolor[HTML]{D9E2F3}                               & \textbf{4}                                                       & -                                                                 & 7680:10240                                                        & 5120:6400                                                         & 3840: 4480                                                      & 5440:5760                                                        & 0:160                                                            & 720:800                                                         & 680:720                                                         \\
\multirow{-10}{*}{\textbf{High Noise}}   & \multirow{-5}{*}{\cellcolor[HTML]{D9E2F3}\textbf{RMS}} & \cellcolor[HTML]{D9E2F3}\textbf{5}                               & \cellcolor[HTML]{D9E2F3}-                                         & \cellcolor[HTML]{D9E2F3}-                                         & \cellcolor[HTML]{D9E2F3}6400:7680                                 & \cellcolor[HTML]{D9E2F3}4480:5120                               & \cellcolor[HTML]{D9E2F3}640: 960                                 & \cellcolor[HTML]{D9E2F3}4160:4320                                & \cellcolor[HTML]{D9E2F3}4240:4320                               & \cellcolor[HTML]{D9E2F3}2600:2640                               \\ \hline
                                         & \cellcolor[HTML]{F2F2F2}                               & \cellcolor[HTML]{F2F2F2}\textbf{1}                               & \cellcolor[HTML]{F2F2F2}-                                         & \cellcolor[HTML]{F2F2F2}-                                         & \cellcolor[HTML]{F2F2F2}-                                         & \cellcolor[HTML]{F2F2F2}-                                       & \cellcolor[HTML]{F2F2F2}-                                        & \cellcolor[HTML]{F2F2F2}-                                        & \cellcolor[HTML]{F2F2F2}-                                       & \cellcolor[HTML]{F2F2F2}-                                       \\
                                         & \cellcolor[HTML]{F2F2F2}                               & \textbf{2}                                                       & -                                                                 & -                                                                 & -                                                                 & -                                                               & -                                                                & -                                                                & -                                                               & -                                                               \\
                                         & \cellcolor[HTML]{F2F2F2}                               & \cellcolor[HTML]{F2F2F2}\textbf{3}                               & \cellcolor[HTML]{F2F2F2}-                                         & \cellcolor[HTML]{F2F2F2}-                                         & \cellcolor[HTML]{F2F2F2}-                                         & \cellcolor[HTML]{F2F2F2}-                                       & \cellcolor[HTML]{F2F2F2}-                                        & \cellcolor[HTML]{F2F2F2}-                                        & \cellcolor[HTML]{F2F2F2}-                                       & \cellcolor[HTML]{F2F2F2}-                                       \\
                                         & \cellcolor[HTML]{F2F2F2}                               & \textbf{4}                                                       & -                                                                 & -                                                                 & -                                                                 & -                                                               & -                                                                & -                                                                & -                                                               & -                                                               \\
                                         & \multirow{-5}{*}{\cellcolor[HTML]{F2F2F2}\textbf{BRF}} & \cellcolor[HTML]{F2F2F2}\textbf{5}                               & \cellcolor[HTML]{F2F2F2}-                                         & \cellcolor[HTML]{F2F2F2}-                                         & \cellcolor[HTML]{F2F2F2}-                                         & \cellcolor[HTML]{F2F2F2}-                                       & \cellcolor[HTML]{F2F2F2}-                                        & \cellcolor[HTML]{F2F2F2}-                                        & \cellcolor[HTML]{F2F2F2}-                                       & \cellcolor[HTML]{F2F2F2}-                                       \\
                                         & \cellcolor[HTML]{D9E2F3}                               & \cellcolor[HTML]{D9E2F3}\textbf{1}                               & \cellcolor[HTML]{D9E2F3}5120:10240                                & \cellcolor[HTML]{D9E2F3}5120:7680                                 & \cellcolor[HTML]{D9E2F3}5120:6400                                 & \cellcolor[HTML]{D9E2F3}0:640                                   & \cellcolor[HTML]{D9E2F3}6080:6400                                & \cellcolor[HTML]{D9E2F3}6240:6400                                & \cellcolor[HTML]{D9E2F3}6320:6400                               & \cellcolor[HTML]{D9E2F3}5680:5720                               \\
                                         & \cellcolor[HTML]{D9E2F3}                               & \textbf{2}                                                       & 0:5120                                                            & 2560:5120                                                         & 0:1280                                                            & 2560:3200                                                       & 2880:3200                                                        & 4880:5040                                                        & 4560:4640                                                       & 4560:4600                                                       \\
                                         & \cellcolor[HTML]{D9E2F3}                               & \cellcolor[HTML]{D9E2F3}\textbf{3}                               & \cellcolor[HTML]{D9E2F3}-                                         & \cellcolor[HTML]{D9E2F3}7680:10240                                & \cellcolor[HTML]{D9E2F3}2560:3840                                 & \cellcolor[HTML]{D9E2F3}5760:6400                               & \cellcolor[HTML]{D9E2F3}4480:4800                                & \cellcolor[HTML]{D9E2F3}0:160                                    & \cellcolor[HTML]{D9E2F3}5680:5760                               & \cellcolor[HTML]{D9E2F3}1400:1440                               \\
                                         & \cellcolor[HTML]{D9E2F3}                               & \textbf{4}                                                       & -                                                                 & 5120:7680                                                         & 8960:10240                                                        & 7040:7680                                                       & 960:1280                                                         & 5600:5760                                                        & 8480:8560                                                       & 6360:6400                                                       \\
\multirow{-10}{*}{\textbf{Mixed Noise}}  & \multirow{-5}{*}{\cellcolor[HTML]{D9E2F3}\textbf{RMS}} & \cellcolor[HTML]{D9E2F3}\textbf{5}                               & \cellcolor[HTML]{D9E2F3}-                                         & \cellcolor[HTML]{D9E2F3}-                                         & \cellcolor[HTML]{D9E2F3}3840:5120                                 & \cellcolor[HTML]{D9E2F3}5120:5760                               & \cellcolor[HTML]{D9E2F3}0:320                                    & \cellcolor[HTML]{D9E2F3}3040:3200                                & \cellcolor[HTML]{D9E2F3}3280:3360                               & \cellcolor[HTML]{D9E2F3}8520:8560                               \\ \hline
\end{tabular}}
\end{table}

Analyzing the signals (low, medium and high noise) it is noted that, as expected, no band above 5120 hz was selected in the top 5 of relevance of the BRF methodology. In this region only noise was inserted without the presence of any harmonic frequency, and therefore should not be selected. On the other hand, several bands above 5120 hz were selected by \textit{rms}.

It can also be noted that in all rankings for signals low, medium and high noise, the bands address some of the harmonic frequencies inserted in the signal: 30, 120, 500, 700, 750, 2300, 2450, 2600, 2700, 2800 and 3450 hz. What does not occur using only \textit{rms}.

From the analysis it can also be verified that when there is a positive signal-to-noise ratio (SNR), that is, more signal than noise in the sample, even with different levels, the BRF is able to identify the relevant frequency bands of the signal.

Finally, analyzing the condition in which the SNR is equal to zero (amount of noise and signal is equal), it is verified that the methodology presents null ranking. That is, in the first analysis performed, the method identifies that it is not possible to separate (for conventional analysis) relevant bands in the signal, since it is completely mixed with noise. Using only the \textit{rms} value, the analyst would be induced to check all bands since there is energy coming from the noise. In an automatic methodology for extracting features for an artificial intelligence method (e.g., machine learning), it would result in the extraction of irrelevant features and consequently a reduction in the model's accuracy rate.

To make a comparison between the 5 most relevant bands obtained in the ranking with BRF and rms, two analyzes were performed. The first, called Values Analysis (VA), calculates the number of times that the same band appeared in the two methodologies without taking into account the order, and the second, called Position Analysis (PA), checks how many times the positions (order) of the selected bands were equal in the ranking, Table~\ref{tab:ranking_comparision_case1}.

\begin{table}[]
\caption{}
\caption*{Ranking comparison - Case 1}
\label{tab:ranking_comparision_case1}
\resizebox{\textwidth}{!}{
\begin{tabular}{@{}ccc|cccccccc@{}}
\toprule
Signal                        & \textit{\begin{tabular}[c]{@{}c@{}}k-level/ \\ Type of Analysis\end{tabular}} & \begin{tabular}[c]{@{}c@{}}0\\ (BW 10240 hz)\end{tabular} & \begin{tabular}[c]{@{}c@{}}1\\ (BW 5120 hz)\end{tabular} & \begin{tabular}[c]{@{}c@{}}2\\ (BW 2560 hz)\end{tabular} & \begin{tabular}[c]{@{}c@{}}3\\ (BW 1280 hz)\end{tabular} & \begin{tabular}[c]{@{}c@{}}4\\ (BW 640hz)\end{tabular} & \begin{tabular}[c]{@{}c@{}}5\\ (BW 320 hz)\end{tabular} & \begin{tabular}[c]{@{}c@{}}6\\ (BW 160 hz)\end{tabular} & \begin{tabular}[c]{@{}c@{}}7\\ (BW 80 hz)\end{tabular} & \begin{tabular}[c]{@{}c@{}}8\\ (BW 40 hz)\end{tabular} \\ \midrule
\multirow{2}{*}{Low Noise}    & VA                                                                            & 100 \%                                                    & 50 \%                                                    & 50 \%                                                    & 60 \%                                                    & 80 \%                                                  & 100 \%                                                  & 80 \%                                                   & 80 \%                                                  & 80 \%                                                  \\
                              & PA                                                                            & 100 \%                                                    & 50 \%                                                    & 50 \%                                                    & 60 \%                                                    & 0 \%                                                   & 60 \%                                                   & 0 \%                                                    & 0 \%                                                   & 20 \%                                                  \\
\multirow{2}{*}{Medium Noise} & VA                                                                            & 100 \%                                                    & 50 \%                                                    & 50 \%                                                    & 60 \%                                                    & 80 \%                                                  & 100 \%                                                  & 80 \%                                                   & 80 \%                                                  & 80 \%                                                  \\
                              & PA                                                                            & 100 \%                                                    & 50 \%                                                    & 50 \%                                                    & 20 \%                                                    & 20 \%                                                  & 0 \%                                                    & 0 \%                                                    & 0 \%                                                   & 0 \%                                                   \\
\multirow{2}{*}{High Noise}   & VA                                                                            & 100 \%                                                    & 50 \%                                                    & 50 \%                                                    & 40 \%                                                    & 40 \%                                                  & 60 \%                                                   & 60 \%                                                   & 20 \%                                                  & 60 \%                                                  \\
                              & PA                                                                            & 100 \%                                                    & 50 \%                                                    & 0 \%                                                     & 40 \%                                                    & 0 \%                                                   & 40 \%                                                   & 0 \%                                                    & 0 \%                                                   & 20 \%                                                  \\
\multirow{2}{*}{Mixed Noise}  & VA                                                                            & 0 \%                                                      & 0 \%                                                     & 0 \%                                                     & 0 \%                                                     & 0 \%                                                   & 0 \%                                                    & 0 \%                                                    & 0 \%                                                   & 0 \%                                                   \\
                              & PA                                                                            & 0 \%                                                      & 0 \%                                                     & 0 \%                                                     & 0 \%                                                     & 0 \%                                                   & 0 \%                                                    & 0 \%                                                    & 0 \%                                                   & 0 \%                                                   \\ \bottomrule
\end{tabular}

}
\end{table}

Analyzing Table~\ref{tab:ranking_comparision_case1}, it can be verified that the signal where the noise is totally mixed, all the values were different, since according to the BRF, no band was considered relevant. For the other signals, it can be noted that the percentage of values that are repeated in both methodologies is similar for conditions: low and medium noise, where the amount of noise is lower, and reduces for high noise condition where the amount of noise is greater than the previous ones. This happens because the BRF is more robust in the presence of noise, being able to better quantify the bands that are relevant for the analysis when there is higher noise. Regarding the ranking positions (order), it can be noted that the methodologies presented different sequences in most cases.

\subsubsection{Case 2: Bearing dataset}

\begin{figure}[ht]
  \subfloat[Normal condition]{
	\begin{minipage}[c][0.65\width]{
	   0.5\textwidth}
	   \centering
       \label{fig:case2_heat_normal}
	   \includegraphics[width=1\textwidth]{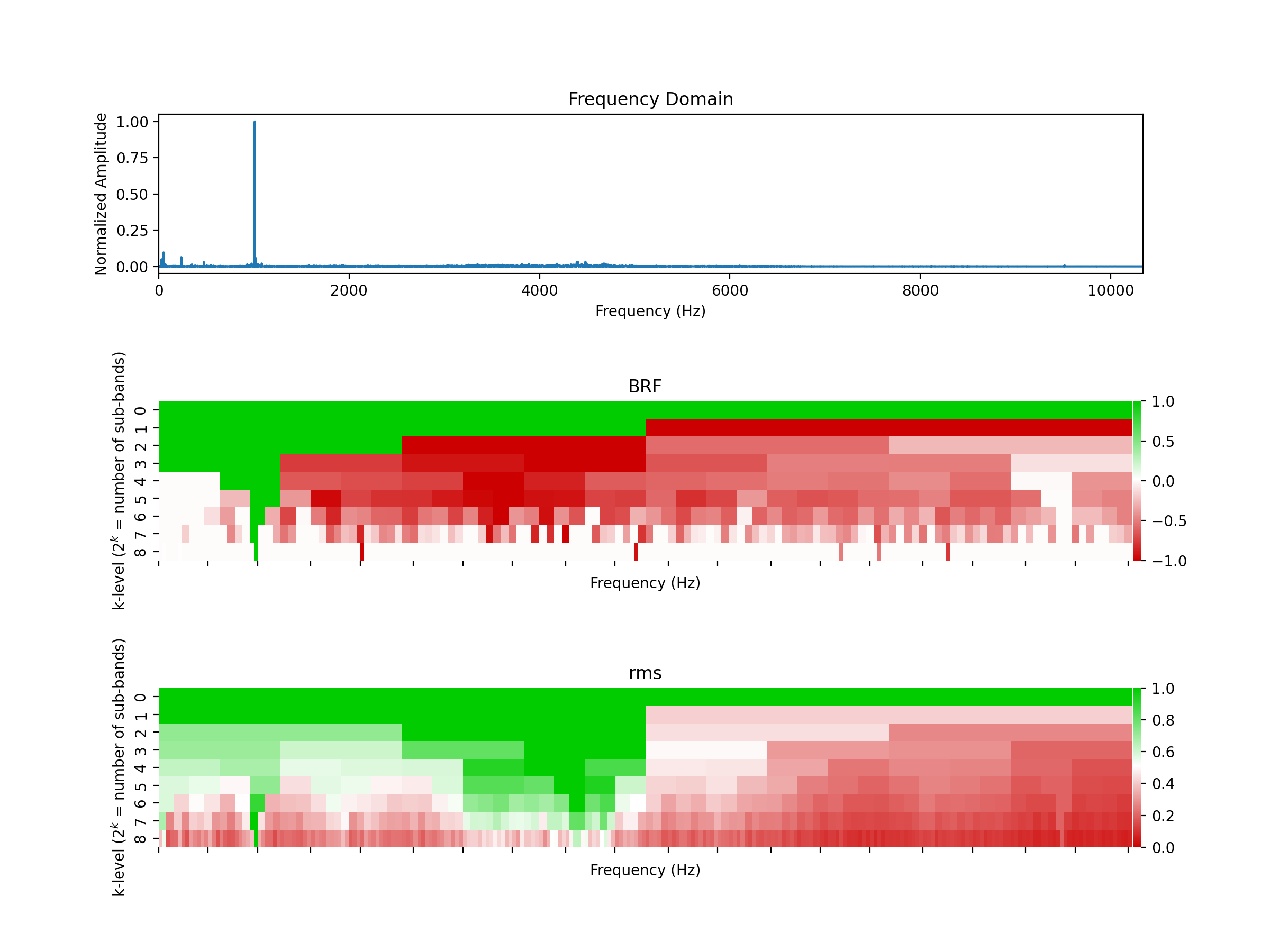}
	\end{minipage}}
 \hfill 
  \subfloat[Fault in outer race]{
	\begin{minipage}[c][0.65\width]{
	   0.5\textwidth}
	   \centering
        \label{fig:case2_heat_fault}
	   \includegraphics[width=1\textwidth]{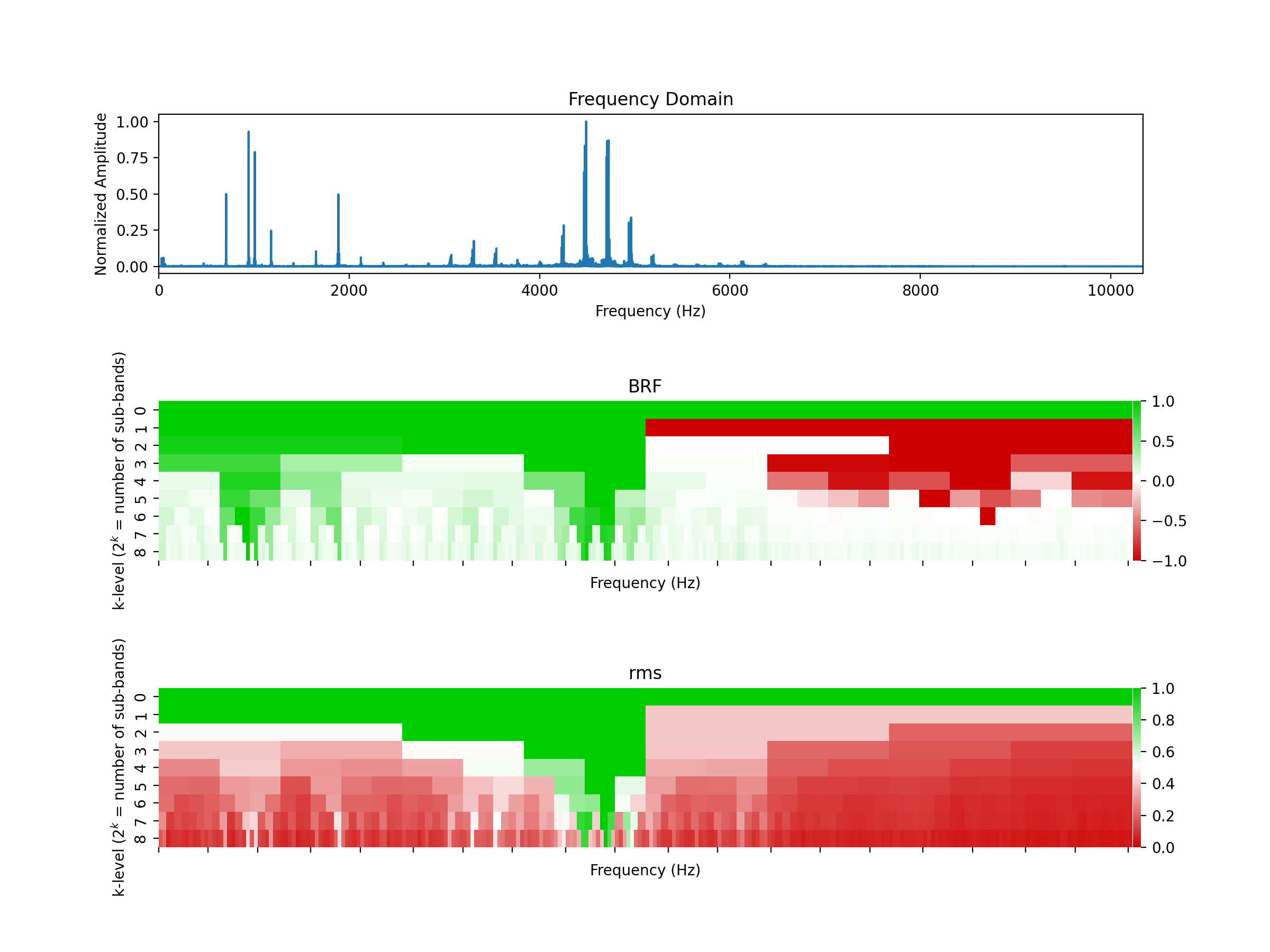}
	\end{minipage}} 
 \hfill 
    \vspace*{-7mm}
    \caption{Heatmap - Case 2.}
    \label{fig:sinalcase3}	
\end{figure}

The heatmap for the two selected signals, normal and fault condition on the outer race, are shown in Fig.\ref{fig:case2_heat_normal} and Fig.\ref{fig:case2_heat_fault}, respectively.

\renewcommand{\baselinestretch}{3} 
\renewcommand{\baselinestretch}{1.5} 

For the normal condition, Fig.\ref{fig:case2_heat_normal}, both methodologies highlighted the band in which the frequency of approximately 1000 Hz was present as the most relevant. In the BRF, only one band was selected as relevant at all levels. However, using the \textit{rms} value, other bands also presented energy close to the maximum value.

In Fig.\ref{fig:case2_heat_fault} where the fault is present, both methodologies identified the high frequency region (excitation of the resonances due to the impact generated by the fault) as relevant. On the other hand, it is known that with the progression of the fault, the excitations present only at high frequency begin to appear in the region of medium and low frequency, associated with the characteristics of the bearing. In this case, the BRF was able to identify not only high frequency excitations as relevant, but also excitations in the region below 2000 Hz. The \textit{rms} value considered only the high frequency region as the most relevant. Knowing that this behavior is associated with the progression of the fault, using only the \textit{rms} the analyst could miss an evolution of the defect, since the low and medium frequencies would not be as prominent as the high ones.

The rankings with the five most relevant bands according to BRF and \textit{rms} for the two signals from Case 2 are shown in Table~\ref{tab:ranking_case2}.

\begin{table}[h!]
\caption{}
\caption*{Relevance Ranking - Case 2}
\label{tab:ranking_case2}
\resizebox{\textwidth}{!}{
\begin{tabular}{@{}cc|ccccccccc@{}}
\hline
\textbf{Signal}                    & \textbf{Feat.}                                         & \textbf{\begin{tabular}[c]{@{}c@{}}Rank/\\ k-level\end{tabular}} & \textbf{\begin{tabular}[c]{@{}c@{}}1\\ (BW 5120 hz)\end{tabular}} & \textbf{\begin{tabular}[c]{@{}c@{}}2\\ (BW 2560 hz)\end{tabular}} & \textbf{\begin{tabular}[c]{@{}c@{}}3\\ (BW 1280 hz)\end{tabular}} & \textbf{\begin{tabular}[c]{@{}c@{}}4\\ (BW 640hz)\end{tabular}} & \textbf{\begin{tabular}[c]{@{}c@{}}5\\ (BW 320 hz)\end{tabular}} & \textbf{\begin{tabular}[c]{@{}c@{}}6\\ (BW 160 hz)\end{tabular}} & \textbf{\begin{tabular}[c]{@{}c@{}}7\\ (BW 80 hz)\end{tabular}} & \textbf{\begin{tabular}[c]{@{}c@{}}8\\ (BW 40 hz)\end{tabular}} \\ \hline
                                   & \cellcolor[HTML]{F2F2F2}                               & \cellcolor[HTML]{F2F2F2}\textbf{1}                               & \cellcolor[HTML]{F2F2F2}0:5120                                    & \cellcolor[HTML]{F2F2F2}0:2560                                    & \cellcolor[HTML]{F2F2F2}0:1280                                    & \cellcolor[HTML]{F2F2F2}640:1280                                & \cellcolor[HTML]{F2F2F2}960:1280                                 & \cellcolor[HTML]{F2F2F2}960:1120                                 & \cellcolor[HTML]{F2F2F2}960:1040                                & \cellcolor[HTML]{F2F2F2}1000:1040                               \\
                                   & \cellcolor[HTML]{F2F2F2}                               & \textbf{2}                                                       & -                                                                 & -                                                                 & -                                                                 & -                                                               & -                                                                & -                                                                & -                                                               & -                                                               \\
                                   & \cellcolor[HTML]{F2F2F2}                               & \cellcolor[HTML]{F2F2F2}\textbf{3}                               & \cellcolor[HTML]{F2F2F2}-                                         & \cellcolor[HTML]{F2F2F2}-                                         & \cellcolor[HTML]{F2F2F2}-                                         & \cellcolor[HTML]{F2F2F2}-                                       & \cellcolor[HTML]{F2F2F2}-                                        & \cellcolor[HTML]{F2F2F2}-                                        & \cellcolor[HTML]{F2F2F2}-                                       & \cellcolor[HTML]{F2F2F2}-                                       \\
                                   & \cellcolor[HTML]{F2F2F2}                               & \textbf{4}                                                       & -                                                                 & -                                                                 & -                                                                 & -                                                               & -                                                                & -                                                                & -                                                               & -                                                               \\
                                   & \multirow{-5}{*}{\cellcolor[HTML]{F2F2F2}\textbf{BRF}} & \cellcolor[HTML]{F2F2F2}\textbf{5}                               & \cellcolor[HTML]{F2F2F2}-                                         & \cellcolor[HTML]{F2F2F2}-                                         & \cellcolor[HTML]{F2F2F2}-                                         & \cellcolor[HTML]{F2F2F2}-                                       & \cellcolor[HTML]{F2F2F2}-                                        & \cellcolor[HTML]{F2F2F2}-                                        & \cellcolor[HTML]{F2F2F2}-                                       & \cellcolor[HTML]{F2F2F2}-                                       \\
                                   & \cellcolor[HTML]{D9E2F3}                               & \cellcolor[HTML]{D9E2F3}\textbf{1}                               & \cellcolor[HTML]{D9E2F3}0:5120                                    & \cellcolor[HTML]{D9E2F3}2560:5120                                 & \cellcolor[HTML]{D9E2F3}3840:5120                                 & \cellcolor[HTML]{D9E2F3}3840: 4480                              & \cellcolor[HTML]{D9E2F3}4160:4480                                & \cellcolor[HTML]{D9E2F3}4320:4480                                & \cellcolor[HTML]{D9E2F3}960:1040                                & \cellcolor[HTML]{D9E2F3}1000:1040                               \\
                                   & \cellcolor[HTML]{D9E2F3}                               & \textbf{2}                                                       & 5120:10240                                                        & 0:2560                                                            & 2560:3840                                                         & 3200:3840                                                       & 4480:4800                                                        & 960:1120                                                         & 4400:4480                                                       & 4400:4440                                                       \\
                                   & \cellcolor[HTML]{D9E2F3}                               & \cellcolor[HTML]{D9E2F3}\textbf{3}                               & \cellcolor[HTML]{D9E2F3}-                                         & \cellcolor[HTML]{D9E2F3}5120:7680                                 & \cellcolor[HTML]{D9E2F3}0:1280                                    & \cellcolor[HTML]{D9E2F3}4480:5120                               & \cellcolor[HTML]{D9E2F3}3520:3840                                & \cellcolor[HTML]{D9E2F3}4640:4800                                & \cellcolor[HTML]{D9E2F3}4320:4400                               & \cellcolor[HTML]{D9E2F3}4360:4400                               \\
                                   & \cellcolor[HTML]{D9E2F3}                               & \textbf{4}                                                       & -                                                                 & 7680:10240                                                        & 1280:2560                                                         & 640:1280                                                        & 3200:3520                                                        & 4480:4640                                                        & 4640:4720                                                       & 4680:4720                                                       \\
\multirow{-10}{*}{\textbf{Normal}} & \multirow{-5}{*}{\cellcolor[HTML]{D9E2F3}\textbf{RMS}} & \cellcolor[HTML]{D9E2F3}\textbf{5}                               & \cellcolor[HTML]{D9E2F3}-                                         & \cellcolor[HTML]{D9E2F3}-                                         & \cellcolor[HTML]{D9E2F3}5120:6400                                 & \cellcolor[HTML]{D9E2F3}0:640                                   & \cellcolor[HTML]{D9E2F3}3840:4160                                & \cellcolor[HTML]{D9E2F3}3520:3680                                & \cellcolor[HTML]{D9E2F3}0:80                                    & \cellcolor[HTML]{D9E2F3}40:80                                   \\ \hline
                                   & \cellcolor[HTML]{F2F2F2}                               & \cellcolor[HTML]{F2F2F2}\textbf{1}                               & \cellcolor[HTML]{F2F2F2}0:5120                                    & \cellcolor[HTML]{F2F2F2}2560:5120                                 & \cellcolor[HTML]{F2F2F2}3840:5120                                 & \cellcolor[HTML]{F2F2F2}4480:5120                               & \cellcolor[HTML]{F2F2F2}4480:4800                                & \cellcolor[HTML]{F2F2F2}800:960                                  & \cellcolor[HTML]{F2F2F2}880:960                                 & \cellcolor[HTML]{F2F2F2}920:960                                 \\
                                   & \cellcolor[HTML]{F2F2F2}                               & \textbf{2}                                                       & -                                                                 & 0:2560                                                            & 0:1280                                                            & 640:1280                                                        & 640:960                                                          & 4640:4800                                                        & 4480:4560                                                       & 4480:4520                                                       \\
                                   & \cellcolor[HTML]{F2F2F2}                               & \cellcolor[HTML]{F2F2F2}\textbf{3}                               & \cellcolor[HTML]{F2F2F2}-                                         & \cellcolor[HTML]{F2F2F2}-                                         & \cellcolor[HTML]{F2F2F2}1280:2560                                 & \cellcolor[HTML]{F2F2F2}3840: 4480                              & \cellcolor[HTML]{F2F2F2}960:1280                                 & \cellcolor[HTML]{F2F2F2}4480:4640                                & \cellcolor[HTML]{F2F2F2}4640:4720                               & \cellcolor[HTML]{F2F2F2}4680:4720                               \\
                                   & \cellcolor[HTML]{F2F2F2}                               & \textbf{4}                                                       & -                                                                 & -                                                                 & -                                                                 & 1280:1920                                                       & 4160:4480                                                        & 960:1120                                                         & 4720:4800                                                       & 4720:4760                                                       \\
                                   & \multirow{-5}{*}{\cellcolor[HTML]{F2F2F2}\textbf{BRF}} & \cellcolor[HTML]{F2F2F2}\textbf{5}                               & \cellcolor[HTML]{F2F2F2}-                                         & \cellcolor[HTML]{F2F2F2}-                                         & \cellcolor[HTML]{F2F2F2}-                                         & \cellcolor[HTML]{F2F2F2}3200:3840                               & \cellcolor[HTML]{F2F2F2}1600:1920                                & \cellcolor[HTML]{F2F2F2}4320:4480                                & \cellcolor[HTML]{F2F2F2}4400:4480                               & \cellcolor[HTML]{F2F2F2}4440:4480                               \\
                                   & \cellcolor[HTML]{D9E2F3}                               & \cellcolor[HTML]{D9E2F3}\textbf{1}                               & \cellcolor[HTML]{D9E2F3}0:5120                                    & \cellcolor[HTML]{D9E2F3}2560:5120                                 & \cellcolor[HTML]{D9E2F3}3840:5120                                 & \cellcolor[HTML]{D9E2F3}4480:5120                               & \cellcolor[HTML]{D9E2F3}4480:4800                                & \cellcolor[HTML]{D9E2F3}4640:4800                                & \cellcolor[HTML]{D9E2F3}4640:4720                               & \cellcolor[HTML]{D9E2F3}4680:4720                               \\
                                   & \cellcolor[HTML]{D9E2F3}                               & \textbf{2}                                                       & 5120:10240                                                        & 0:2560                                                            & 2560:3840                                                         & 3840: 4480                                                      & 4160:4480                                                        & 4480:4640                                                        & 4480:4560                                                       & 4480:4520                                                       \\
                                   & \cellcolor[HTML]{D9E2F3}                               & \cellcolor[HTML]{D9E2F3}\textbf{3}                               & \cellcolor[HTML]{D9E2F3}-                                         & \cellcolor[HTML]{D9E2F3}5120:7680                                 & \cellcolor[HTML]{D9E2F3}0:1280                                    & \cellcolor[HTML]{D9E2F3}3200:3840                               & \cellcolor[HTML]{D9E2F3}4800:5120                                & \cellcolor[HTML]{D9E2F3}4320:4480                                & \cellcolor[HTML]{D9E2F3}4400:4480                               & \cellcolor[HTML]{D9E2F3}4440:4480                               \\
                                   & \cellcolor[HTML]{D9E2F3}                               & \textbf{4}                                                       & -                                                                 & 7680:10240                                                        & 5120:6400                                                         & 640:1280                                                        & 3520:3840                                                        & 4160:4320                                                        & 4720:4800                                                       & 4720:4760                                                       \\
\multirow{-10}{*}{\textbf{Fault}}  & \multirow{-5}{*}{\cellcolor[HTML]{D9E2F3}\textbf{RMS}} & \cellcolor[HTML]{D9E2F3}\textbf{5}                               & \cellcolor[HTML]{D9E2F3}-                                         & \cellcolor[HTML]{D9E2F3}-                                         & \cellcolor[HTML]{D9E2F3}1280:2560                                 & \cellcolor[HTML]{D9E2F3}5120:5760                               & \cellcolor[HTML]{D9E2F3}3200:3520                                & \cellcolor[HTML]{D9E2F3}4800:4960                                & \cellcolor[HTML]{D9E2F3}4880:4960                               & \cellcolor[HTML]{D9E2F3}4920:4960                               \\ \hline

\end{tabular}}
\end{table}

As in Case 1, the ranking of the BRF confirms the selection of bands related to the most relevant frequencies in the signal, associated with the fault diagnosis. For the normal situation, only one band was selected as relevant, and for the fault condition, different bands in low, medium and high frequencies were selected, according to the fault progression characteristic mentioned above. On the other hand, the \textit{rms} value no longer points to frequencies that are directly associated with the defect present in the signal.

The comparison between the top 5 values of the relevance rankings obtained with BRF and \textit{rms} is presented in Table~\ref{tab:ranking_comparision_case2}.

\begin{table}[]
\caption{}
\caption*{Ranking comparison - Case 2}
\label{tab:ranking_comparision_case2}
\resizebox{\textwidth}{!}{
\begin{tabular}{@{}ccccccccccc@{}}
\toprule
Signal                  & \textit{\begin{tabular}[c]{@{}c@{}}k-level/ \\ Type of Analysis\end{tabular}} & \begin{tabular}[c]{@{}c@{}}0\\ (BW 10240 hz)\end{tabular} & \begin{tabular}[c]{@{}c@{}}1\\ (BW 5120 hz)\end{tabular} & \begin{tabular}[c]{@{}c@{}}2\\ (BW 2560 hz)\end{tabular} & \begin{tabular}[c]{@{}c@{}}3\\ (BW 1280 hz)\end{tabular} & \begin{tabular}[c]{@{}c@{}}4\\ (BW 640hz)\end{tabular} & \begin{tabular}[c]{@{}c@{}}5\\ (BW 320 hz)\end{tabular} & \begin{tabular}[c]{@{}c@{}}6\\ (BW 160 hz)\end{tabular} & \begin{tabular}[c]{@{}c@{}}7\\ (BW 80 hz)\end{tabular} & \begin{tabular}[c]{@{}c@{}}8\\ (BW 40 hz)\end{tabular} \\ \midrule
\multirow{2}{*}{Normal} & VA                                                                            & 100 \%                                                    & 50 \%                                                    & 25 \%                                                    & 20 \%                                                    & 40 \%                                                  & 0 \%                                                    & 20 \%                                                   & 40 \%                                                  & 40 \%                                                  \\
                        & PA                                                                            & 100 \%                                                    & 50 \%                                                    & 0 \%                                                     & 0 \%                                                     & 0 \%                                                   & 0 \%                                                    & 0 \%                                                    & 20 \%                                                  & 20 \%                                                  \\
\multirow{2}{*}{Fault}  & VA                                                                            & 100 \%                                                    & 50 \%                                                    & 50 \%                                                    & 60 \%                                                    & 80 \%                                                  & 40 \%                                                   & 60 \%                                                   & 80 \%                                                  & 80 \%                                                  \\
                        & PA                                                                            & 100 \%                                                    & 50 \%                                                    & 50 \%                                                    & 20 \%                                                    & 20 \%                                                  & 20 \%                                                   & 0 \%                                                    & 40 \%                                                  & 40 \%                                                  \\ \bottomrule

\end{tabular}}
\end{table}

It can be noted that in the normal condition, the methodologies presented more different values, since the \textit{rms} considered the high frequency region relevant in this situation. On the other hand, the fault situation presented some similar values (mainly related to high frequency), and the main difference is related to the low frequency region considered by BRF and not by the \textit{rms}.

\subsubsection{Case 3: Mechanical fault dataset}

\begin{figure}[ht]
  \subfloat[Normal condition]{
	\begin{minipage}[c][0.65\width]{
	   0.5\textwidth}
	   \centering
       \label{fig:case3_heat_normal}
	   \includegraphics[width=1\textwidth]{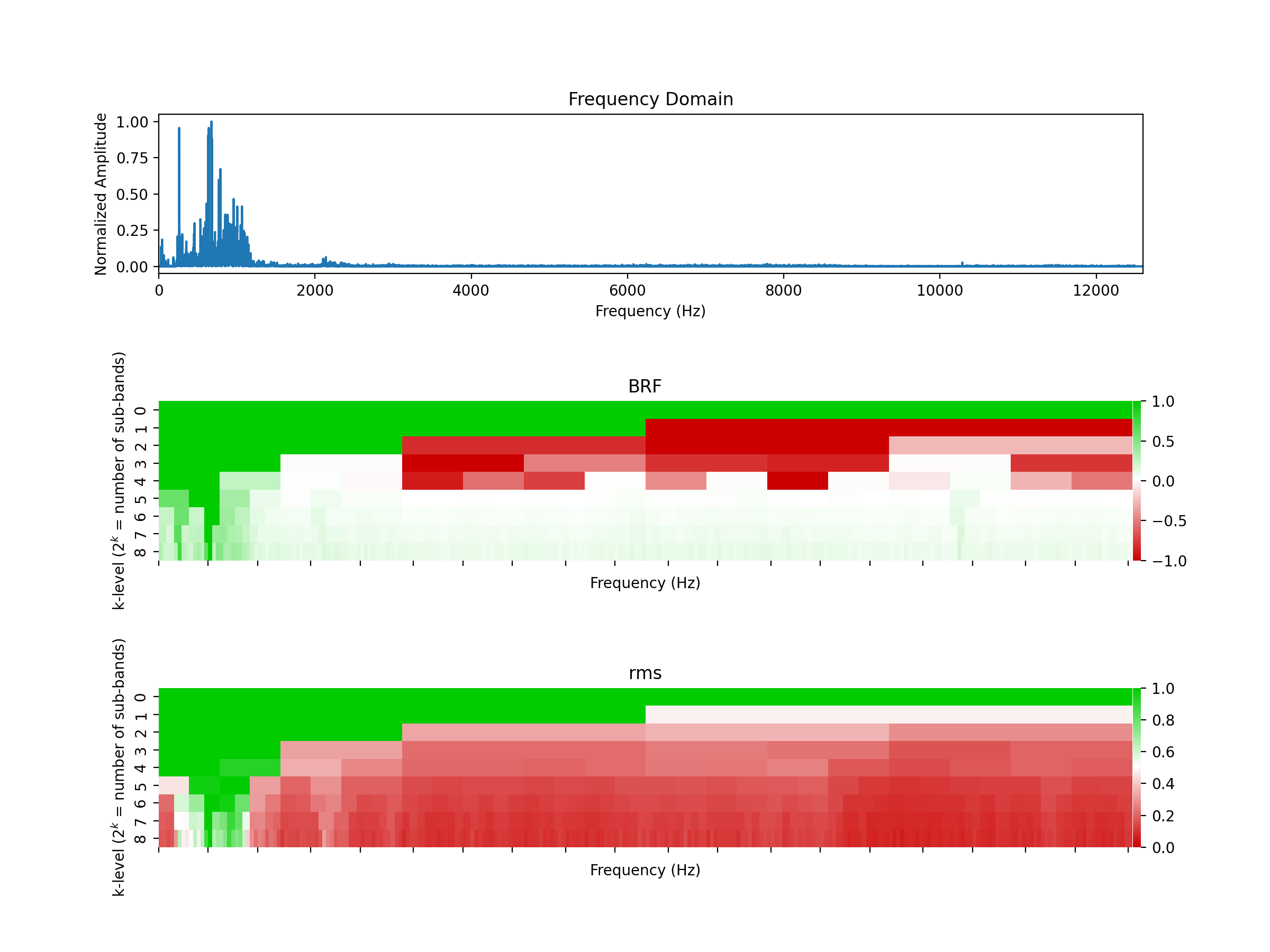}
	\end{minipage}}
 \hfill 
  \subfloat[Unbalance]{
	\begin{minipage}[c][0.65\width]{
	   0.5\textwidth}
	   \centering
        \label{fig:case3_heat_fault}
	   \includegraphics[width=1\textwidth]{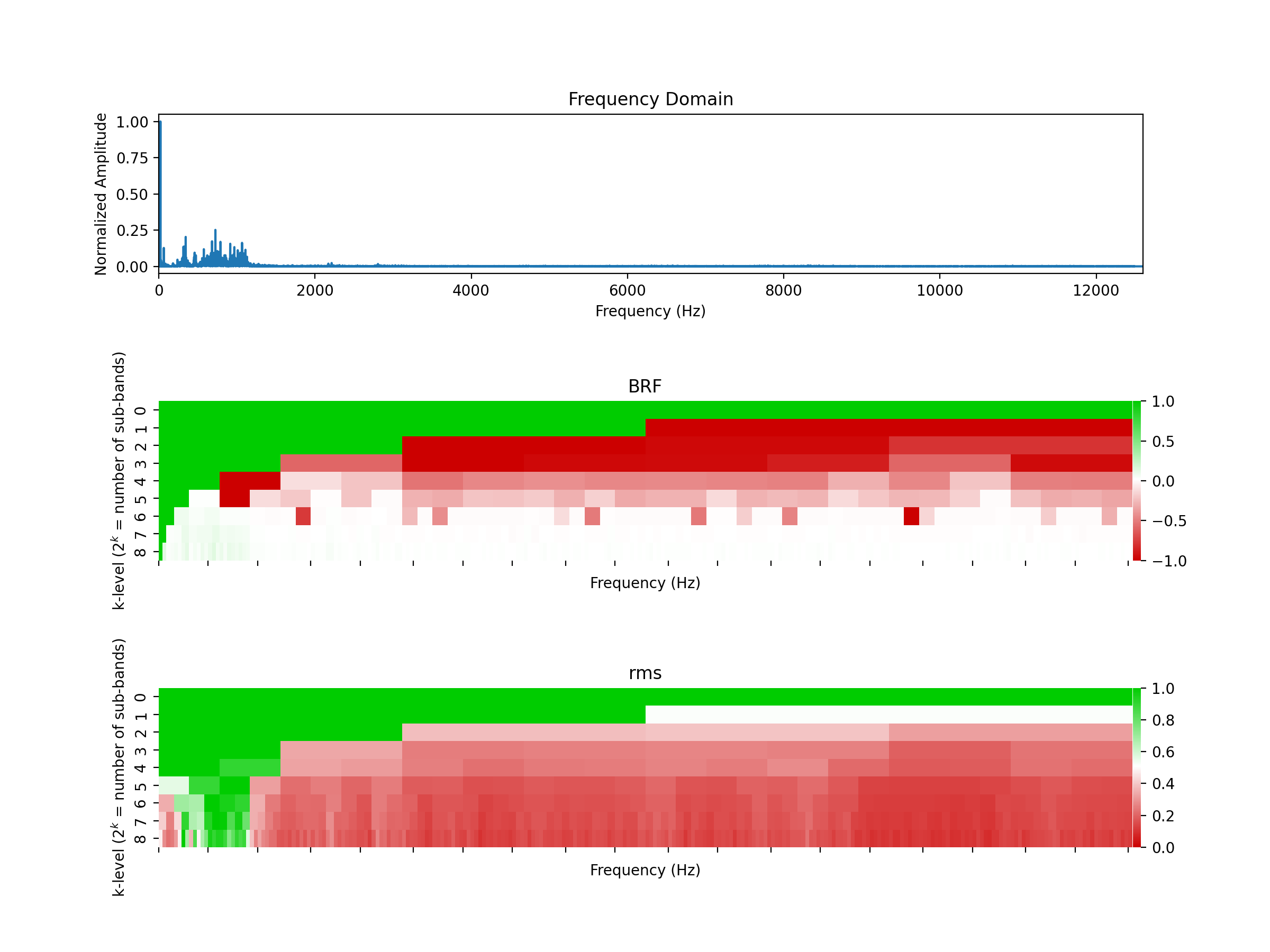}
	\end{minipage}} 
 \hfill 
    \vspace*{-7mm}
    \caption{Heatmap - Case 3.}
    \label{fig:sinalcase3}	
\end{figure}

The heatmap for the two selected signals, normal and unbalance, are shown in Fig.\ref{fig:case3_heat_normal} and Fig.\ref{fig:case3_heat_fault}, respectively.

\renewcommand{\baselinestretch}{3} 
\renewcommand{\baselinestretch}{1.5} 

For the normal condition, Fig.\ref{fig:case3_heat_normal} both methodologies selected similar frequency bands, mainly at levels 5, 6 and 7. For the fault condition, Fig.\ref{fig:case3_heat_fault}, as it is an unbalance, the most relevant frequency in the signal is related to the rotation frequency, which in the case it is at 21.9 hz. Analyzing the result obtained by the BRF, it is noted that the band that contained the frequency was selected as the most relevant in each of the levels, validating the methodology. On the other hand, using the \textit{rms}, the region presented lower or no relevance when analyzing each level.

The rankings with the five most relevant bands according to BRF and \textit{rms} for the two signals from Case 3 are shown in Table~\ref{tab:ranking_case3}.

\begin{table}[]
\caption{}
\caption*{Relevance Ranking - Case 3}
\label{tab:ranking_case3}
\resizebox{\textwidth}{!}{
\begin{tabular}{@{}cc|ccccccccc@{}}

\hline
\textbf{Signal}                    & \textbf{Feat.}                                         & \textbf{\begin{tabular}[c]{@{}c@{}}Rank/\\ k-level\end{tabular}} & \textbf{\begin{tabular}[c]{@{}c@{}}1\\ (BW 6250 hz)\end{tabular}} & \textbf{\begin{tabular}[c]{@{}c@{}}2\\ (BW 3125 hz)\end{tabular}} & \textbf{\begin{tabular}[c]{@{}c@{}}3\\ (BW 1562 hz)\end{tabular}} & \textbf{\begin{tabular}[c]{@{}c@{}}4\\ (BW 781 hz)\end{tabular}} & \textbf{\begin{tabular}[c]{@{}c@{}}5\\ (BW 390 hz)\end{tabular}} & \textbf{\begin{tabular}[c]{@{}c@{}}6\\ (BW 195 hz)\end{tabular}} & \textbf{\begin{tabular}[c]{@{}c@{}}7\\ (BW 97 hz)\end{tabular}} & \textbf{\begin{tabular}[c]{@{}c@{}}8\\ (BW 48 hz)\end{tabular}} \\ \hline
                                   & \cellcolor[HTML]{F2F2F2}                               & \cellcolor[HTML]{F2F2F2}\textbf{1}                               & \cellcolor[HTML]{F2F2F2}0:6250                                    & \cellcolor[HTML]{F2F2F2}0:3125                                    & \cellcolor[HTML]{F2F2F2}0:1562                                    & \cellcolor[HTML]{F2F2F2}0:781                                    & \cellcolor[HTML]{F2F2F2}390:781                                  & \cellcolor[HTML]{F2F2F2}585:781                                  & \cellcolor[HTML]{F2F2F2}585:683                                 & \cellcolor[HTML]{F2F2F2}634:683                                 \\
                                   & \cellcolor[HTML]{F2F2F2}                               & \textbf{2}                                                       & -                                                                 & -                                                                 & -                                                                 & 781:1562                                                         & 0:390                                                            & 195:390                                                          & 195:293                                                         & 244:292                                                         \\
                                   & \cellcolor[HTML]{F2F2F2}                               & \cellcolor[HTML]{F2F2F2}\textbf{3}                               & \cellcolor[HTML]{F2F2F2}-                                         & \cellcolor[HTML]{F2F2F2}-                                         & \cellcolor[HTML]{F2F2F2}-                                         & \cellcolor[HTML]{F2F2F2}-                                        & \cellcolor[HTML]{F2F2F2}781:1171                                 & \cellcolor[HTML]{F2F2F2}781:976                                  & \cellcolor[HTML]{F2F2F2}781:878                                 & \cellcolor[HTML]{F2F2F2}585:633                                 \\
                                   & \cellcolor[HTML]{F2F2F2}                               & \textbf{4}                                                       & -                                                                 & -                                                                 & -                                                                 & -                                                                & 1171:1561                                                        & 976:1171                                                         & 683:781                                                         & 732:812                                                         \\
                                   & \multirow{-5}{*}{\cellcolor[HTML]{F2F2F2}\textbf{BRF}} & \cellcolor[HTML]{F2F2F2}\textbf{5}                               & \cellcolor[HTML]{F2F2F2}-                                         & \cellcolor[HTML]{F2F2F2}-                                         & \cellcolor[HTML]{F2F2F2}-                                         & \cellcolor[HTML]{F2F2F2}-                                        & \cellcolor[HTML]{F2F2F2}10156:10546                              & \cellcolor[HTML]{F2F2F2}0:195                                    & \cellcolor[HTML]{F2F2F2}878:976                                 & \cellcolor[HTML]{F2F2F2}781:829                                 \\
                                   & \cellcolor[HTML]{D9E2F3}                               & \cellcolor[HTML]{D9E2F3}\textbf{1}                               & \cellcolor[HTML]{D9E2F3}0:6250                                    & \cellcolor[HTML]{D9E2F3}0:3125                                    & \cellcolor[HTML]{D9E2F3}0:1562                                    & \cellcolor[HTML]{D9E2F3}0:781                                    & \cellcolor[HTML]{D9E2F3}781:1171                                 & \cellcolor[HTML]{D9E2F3}585:781                                  & \cellcolor[HTML]{D9E2F3}585:683                                 & \cellcolor[HTML]{D9E2F3}634:683                                 \\
                                   & \cellcolor[HTML]{D9E2F3}                               & \textbf{2}                                                       & 6250:12500                                                        & 6250:9375                                                         & 1562:3124                                                         & 781:1562                                                         & 390:781                                                          & 781:976                                                          & 878:976                                                         & 878:927                                                         \\
                                   & \cellcolor[HTML]{D9E2F3}                               & \cellcolor[HTML]{D9E2F3}\textbf{3}                               & \cellcolor[HTML]{D9E2F3}                                          & \cellcolor[HTML]{D9E2F3}3125:6250                                 & \cellcolor[HTML]{D9E2F3}6250:7812                                 & \cellcolor[HTML]{D9E2F3}1562:2343                                & \cellcolor[HTML]{D9E2F3}0:390                                    & \cellcolor[HTML]{D9E2F3}976:1171                                 & \cellcolor[HTML]{D9E2F3}976:1074                                & \cellcolor[HTML]{D9E2F3}585:633                                 \\
                                   & \cellcolor[HTML]{D9E2F3}                               & \textbf{4}                                                       &                                                                   & 9375:12500                                                        & 7812:9375                                                         & 2343:3125                                                        & 1171:1561                                                        & 390:585                                                          & 781:878                                                         & 927:976                                                         \\
\multirow{-10}{*}{\textbf{Normal}} & \multirow{-5}{*}{\cellcolor[HTML]{D9E2F3}\textbf{RMS}} & \cellcolor[HTML]{D9E2F3}\textbf{5}                               & \cellcolor[HTML]{D9E2F3}                                          & \cellcolor[HTML]{D9E2F3}                                          & \cellcolor[HTML]{D9E2F3}3125:4687                                 & \cellcolor[HTML]{D9E2F3}7812:8593                                & \cellcolor[HTML]{D9E2F3}1953:2343                                & \cellcolor[HTML]{D9E2F3}195:390                                  & \cellcolor[HTML]{D9E2F3}683:781                                 & \cellcolor[HTML]{D9E2F3}1025:1074                               \\ \hline
                                   & \cellcolor[HTML]{F2F2F2}                               & \cellcolor[HTML]{F2F2F2}\textbf{1}                               & \cellcolor[HTML]{F2F2F2}0:6250                                    & \cellcolor[HTML]{F2F2F2}0:3125                                    & \cellcolor[HTML]{F2F2F2}0:1562                                    & \cellcolor[HTML]{F2F2F2}0:781                                    & \cellcolor[HTML]{F2F2F2}0:390                                    & \cellcolor[HTML]{F2F2F2}0:195                                    & \cellcolor[HTML]{F2F2F2}0:97                                    & \cellcolor[HTML]{F2F2F2}0:48                                    \\
                                   & \cellcolor[HTML]{F2F2F2}                               & \textbf{2}                                                       & -                                                                 & -                                                                 & -                                                                 & -                                                                & -                                                                & 195:390                                                          & 683:780                                                         & 683:731                                                         \\
                                   & \cellcolor[HTML]{F2F2F2}                               & \cellcolor[HTML]{F2F2F2}\textbf{3}                               & \cellcolor[HTML]{F2F2F2}-                                         & \cellcolor[HTML]{F2F2F2}-                                         & \cellcolor[HTML]{F2F2F2}-                                         & \cellcolor[HTML]{F2F2F2}-                                        & \cellcolor[HTML]{F2F2F2}-                                        & \cellcolor[HTML]{F2F2F2}-                                        & \cellcolor[HTML]{F2F2F2}293:390                                 & \cellcolor[HTML]{F2F2F2}293:341                                 \\
                                   & \cellcolor[HTML]{F2F2F2}                               & \textbf{4}                                                       & -                                                                 & -                                                                 & -                                                                 & -                                                                & -                                                                & -                                                                & -                                                               & 878:927                                                         \\
                                   & \multirow{-5}{*}{\cellcolor[HTML]{F2F2F2}\textbf{BRF}} & \cellcolor[HTML]{F2F2F2}\textbf{5}                               & \cellcolor[HTML]{F2F2F2}-                                         & \cellcolor[HTML]{F2F2F2}-                                         & \cellcolor[HTML]{F2F2F2}-                                         & \cellcolor[HTML]{F2F2F2}-                                        & \cellcolor[HTML]{F2F2F2}-                                        & \cellcolor[HTML]{F2F2F2}-                                        & \cellcolor[HTML]{F2F2F2}-                                       & \cellcolor[HTML]{F2F2F2}585:633                                 \\
                                   & \cellcolor[HTML]{D9E2F3}                               & \cellcolor[HTML]{D9E2F3}\textbf{1}                               & \cellcolor[HTML]{D9E2F3}0:6250                                    & \cellcolor[HTML]{D9E2F3}0:3125                                    & \cellcolor[HTML]{D9E2F3}0:1562                                    & \cellcolor[HTML]{D9E2F3}0:781                                    & \cellcolor[HTML]{D9E2F3}781:1171                                 & \cellcolor[HTML]{D9E2F3}585:781                                  & \cellcolor[HTML]{D9E2F3}683:781                                 & \cellcolor[HTML]{D9E2F3}293:341                                 \\
                                   & \cellcolor[HTML]{D9E2F3}                               & \textbf{2}                                                       & 6250:12500                                                        & 6250:9375                                                         & 1562:3124                                                         & 781:1562                                                         & 390:781                                                          & 781:976                                                          & 781:878                                                         & 634:683                                                         \\
                                   & \cellcolor[HTML]{D9E2F3}                               & \cellcolor[HTML]{D9E2F3}\textbf{3}                               & \cellcolor[HTML]{D9E2F3}-                                         & \cellcolor[HTML]{D9E2F3}3125:6250                                 & \cellcolor[HTML]{D9E2F3}6250:7812                                 & \cellcolor[HTML]{D9E2F3}1562:2343                                & \cellcolor[HTML]{D9E2F3}0:390                                    & \cellcolor[HTML]{D9E2F3}976:1171                                 & \cellcolor[HTML]{D9E2F3}976:1074                                & \cellcolor[HTML]{D9E2F3}731:781                                 \\
                                   & \cellcolor[HTML]{D9E2F3}                               & \textbf{4}                                                       & -                                                                 & 9375:12500                                                        & 4687: 6250                                                        & 2343:3125                                                        & 1171:1561                                                        & 195:390                                                          & 585:683                                                         & 781:829                                                         \\
\multirow{-10}{*}{\textbf{Fault}}  & \multirow{-5}{*}{\cellcolor[HTML]{D9E2F3}\textbf{RMS}} & \cellcolor[HTML]{D9E2F3}\textbf{5}                               & \cellcolor[HTML]{D9E2F3}-                                         & \cellcolor[HTML]{D9E2F3}-                                         & \cellcolor[HTML]{D9E2F3}7812:9375                                 & \cellcolor[HTML]{D9E2F3}7812:8593                                & \cellcolor[HTML]{D9E2F3}1953:2343                                & \cellcolor[HTML]{D9E2F3}390:585                                  & \cellcolor[HTML]{D9E2F3}293:390                                 & \cellcolor[HTML]{D9E2F3}976:1024                                \\ \hline

\end{tabular}}
\end{table}

The ranking of the BRF for the unbalance condition confirms the presence of the rotation frequency (associated with fault) in the selected band as the most relevant in each of the levels.

The comparison between the top 5 values of the relevance rankings obtained with BRF and \textit{rms} is presented in Table~\ref{tab:ranking_comparision_case3}.

\begin{table}[]
\caption{}
\caption*{Ranking comparison - Case 3}
\label{tab:ranking_comparision_case3}
\resizebox{\textwidth}{!}{
\begin{tabular}{@{}ccccccccccc@{}}
\toprule
Signal                  & \textit{\begin{tabular}[c]{@{}c@{}}k-level/ \\ Type of Analysis\end{tabular}} & \begin{tabular}[c]{@{}c@{}}0\\ (BW 12500 hz)\end{tabular} & \begin{tabular}[c]{@{}c@{}}1\\ (BW 6250 hz)\end{tabular} & \begin{tabular}[c]{@{}c@{}}2\\ (BW 3125 hz)\end{tabular} & \begin{tabular}[c]{@{}c@{}}3\\ (BW 1562 hz)\end{tabular} & \begin{tabular}[c]{@{}c@{}}4\\ (BW 781 hz)\end{tabular} & \begin{tabular}[c]{@{}c@{}}5\\ (BW 390 hz)\end{tabular} & \begin{tabular}[c]{@{}c@{}}6\\ (BW 195 hz)\end{tabular} & \begin{tabular}[c]{@{}c@{}}7\\ (BW 97 hz)\end{tabular} & \begin{tabular}[c]{@{}c@{}}8\\ (BW 48 hz)\end{tabular} \\ \midrule
\multirow{2}{*}{Normal} & VA                                                                            & 100 \%                                                    & 50 \%                                                    & 25 \%                                                    & 20 \%                                                    & 40 \%                                                   & 80 \%                                                   & 80 \%                                                   & 80 \%                                                  & 40 \%                                                  \\
                        & PA                                                                            & 100 \%                                                    & 50 \%                                                    & 25 \%                                                    & 20 \%                                                    & 40 \%                                                   & 20 \%                                                   & 20 \%                                                   & 20 \%                                                  & 40 \%                                                  \\
\multirow{2}{*}{Fault}  & VA                                                                            & 100 \%                                                    & 50 \%                                                    & 25 \%                                                    & 20 \%                                                    & 20 \%                                                   & 20 \%                                                   & 20 \%                                                   & 40 \%                                                  & 0 \%                                                   \\
                        & PA                                                                            & 100 \%                                                    & 50 \%                                                    & 25 \%                                                    & 20 \%                                                    & 20 \%                                                   & 0 \%                                                    & 0 \%                                                    & 0 \%                                                   & 0 \%                                                   \\ \bottomrule
\end{tabular}}
\end{table}

It can be noted that there are differences in the values and positions selected in the top 5 of each methodology, with greater similarity for levels 5, 6 and 7 in relation to values (normal condition, mainly). As in the other Cases, the positions of the rankings present greater differences.

\section{Conclusions}

A new approach that allows to automatically select the relevant frequency bands for an analysis of rotating machinery signals, and to obtain the respective ranking of importance is presented. Through the analysis of the entropy and \textit{rms} value of the signal, the BRF is obtained allowing the classification of relevance of the band. In automatic systems (e.g., Artificial Intelligence frameworks) the use of the method allows analyzing only bands that contain relevant information, and in manual analyses, it provides the vibration specialist with guidance on the main focuses of the analysis.

The results obtained for the different dataset show that the BRF is able to identify the bands that present relevant information for the analysis of rotating machinery. The methodology was able to identify the relevance of bands both for synthetically generated data, with and without the presence of noise, and for two real dataset.

Due to its unsupervised characteristic, the methodology can be applied within frameworks as a pre-feature extraction method in data analytics and artificial intelligence applications, avoiding extracting features from irrelevant bands of the signal. The BRF also makes a brief contribution to the development of XAI (when applied together with AI techniques), by providing the ranking of relevance of the bands. It can also be used to verify the correct operation of sensors (quality of the acquired signal), contributing to the reliability of wireless/remote and IoT (Internet of Things) monitoring systems.

Future works will explore the possibility of using the BRF value as an anomaly detection / drift concept detection method. Considering the variation in the classification of a band from relevant to irrelevant during a time series may indicate variations or anomalies in the system. Or suggesting that the model needs to be retrained due to a new distribution of the data.

In addition, the BRF will be studied as a monitoring feature over time (trend analysis) in order to assess its sensitivity as an indicator of the component's end-of-life. Future developments also include studying the behavior of the method in time series of different applications (e.g., electroencephalogram (EEG) etc).

\section*{Acknowledgement}
\addcontentsline{toc}{section}{Acknowledgement}

The authors gratefully acknowledge the Brazilian research funding agencies CNPq (National Council for Scientific and Technological Development) and CAPES (Federal Agency for the Support and Improvement of Higher Education) for their financial support of this work. The Italian Government PNRR iniatiatives 'Partenariato 11: Made in Italy circolare e sostenibile' and 'Ecosistema dell'Innovazione  - iNest' are also gratefully acknowledged.

\addcontentsline{toc}{section}{References}
\bibliography{mybibfile}

\end{document}